%% file: BDX-PAC45-upd.tex
\documentclass[letterpaper]{article}
\usepackage{color,amsmath}
\usepackage{verbatim}
\pdfoutput=1
\usepackage{multirow} 
\usepackage{draftwatermark}
\SetWatermarkLightness{1.0} 
\SetWatermarkScale{7.5}

\newif\ifhyprf
\hyprftrue
    
\RequirePackage{ifpdf}

\ifpdf
    \pdfoutput=1        
    \pdftrue
\else
    \pdffalse           
\fi

\ifpdf
  \definecolor{rltred}{rgb}{0.75,0,0}
  \definecolor{rltgreen}{rgb}{0,0.3,0}
  \definecolor{rltblue}{rgb}{0,0,0.75}
  \definecolor{rltdarkgreen}{rgb}{0.1,0.6,0.1}
  \ifhyprf
     \usepackage[pdftex,
         colorlinks=true,
         urlcolor=rltblue,       
         filecolor=rltgreen,     
         linkcolor=rltred,       
         citecolor=rltdarkgreen, 
         pagebackref,
         pdfpagemode=None,
         pdftitle={Dark matter search in a Beam-Dump eXperiment (BDX) at Jefferson Lab: an update on PR12-16-001},
         pdfauthor={many},
         pdfsubject={Beam dump experiment JLab Hall A},
         pdfkeywords={Hall A, Beam dump, dark photon}]{hyperref}
  \fi
  \usepackage{pdfcolmk}
  \DeclareGraphicsExtensions{.pdf,.png,.jpg}
\else
  \usepackage{graphicx}
  \DeclareGraphicsExtensions{.eps,.epsi,.ps,.eps.gz,.epsi.gz,.ps.gz}
\fi

\usepackage{graphicx,lscape,rotating}
\usepackage{amsmath}
\usepackage{amssymb,amsfonts,textcomp}

\usepackage{lineno}

\input{newcomm.tex}

\usepackage{alphalph}
\makeatletter
\newcommand*{\fnsymbolsingle}[1]{%
\ensuremath{%
\ifcase#1%
\or *%
\or \dagger
\or \ddagger
\or \mathsection
\or \mathparagraph
\else
\@ctrerr \fi
}%
}
\makeatother
\newalphalph{\fnsymbolmult}[mult]{\fnsymbolsingle}{}

\begin{document}
\begin{center}
{\tiny \leftline{V2.2}}
{\tiny\leftline{\thedate}}
\date{\today}
\rightline{PR12-16-001 update to PAC 45}
\rightline{\mbox{FERMILAB-TM-2667-PPD}}
\vskip 1.0cm
{\bf\huge Dark matter search in a Beam-Dump eXperiment (BDX) at Jefferson Lab
\\ an update on PR12-16-001}

\vskip 0.5cm
{  \large \it The BDX Collaboration }

\vskip 0.5cm
{M.~Battaglieri\footnote{Contact Person, email: Marco.Battaglieri@ge.infn.it}\footnote{Spokesperson}, A.~Bersani, G.~Bracco, B.~Caiffi, A.~Celentano$^\dag$, R.~De~Vita$^\dag$,   L.~Marsicano, P.~Musico, M.~Osipenko, F.~Panza
% ...
\\}
{\it\small\msu}
\bigskip 

{K.~Hicks\\}
{\it\small\ohio}
\bigskip

{P.~Cole\\}
{\it\small\idaho}
\bigskip

{S.~Dobbs\\}
{\it\small\northw}
\bigskip

{ C.~Fanelli, P.~T.~M.~Murthy\\}
{\it\small\mito}
\bigskip

\newpage
\begin{abstract}

\textcolor{red} {}
This document is an update to the proposal {\it PR12-16-001 Dark matter search in a Beam-Dump eXperiment (BDX) at Jefferson Lab}~\cite{bdx-proposal}  submitted to JLab-PAC44 in 2016 reporting progress in addressing questions
 raised regarding the beam-on backgrounds. The concerns are addressed by adopting a new simulation tool, FLUKA, and planning measurements of muon fluxes from the dump with its existing shielding around the dump. 
First, we have implemented the detailed BDX experimental geometry into a FLUKA simulation, in consultation with
experts from the JLab Radiation Control Group. The FLUKA simulation has been compared directly to our GEANT4 
simulations and shown to agree in regions of validity. The FLUKA interaction package, with a tuned set of biasing weights, 
is naturally able to generate reliable particle distributions with very small probabilities and therefore predict rates at the 
detector location beyond the planned shielding around the beam dump.  Second, we have developed a plan to 
conduct measurements of the muon flux
from the Hall-A dump in its current configuration to validate our simulations.

\end{abstract}

\vskip 1.0cm
 
\end{center} 

\newpage
\tableofcontents
\newpage


\input{BDX-PAC45-upd-intro}

\input{BDX-PAC45-upd-th}

\input{BDX-PAC45-upd-sim}

\input{BDX-PAC45-upd-mu}

\section{Summary}
In this update we address the issues raised by PAC44 about the assessment of the  beam-on backgrounds in the BDX experiment.
We show that with the use of a new simulation tool, FLUKA, it is possible to simulate a number of electron/beam-dump interactions in the range of the 10$^{22}$ EOT expected in BDX. We used FLUKA to estimate the beam-on background presented  in  PR12-16-001 finding good consistency and confirming previous estimates based on GEANT4.
For an experimental validation of the MC tools we propose a dedicated test to measure the muon flux produced in the Hall-A beam dump and propagating to the  proposed  location of the BDX detector. This test  will  be performed in the current shielding configuration using the same detector technology proposed for BDX. The quantitative comparison of  $\mu$ rates measured in a realistic beam-on background will validate the simulation tools and the proposed detector technology for the BDX experiment.

\clearpage
\clearpage

\newpage

\bibliographystyle{unsrt}                                                                              
\bibliography{BDX-PAC45-bib}

\end{document}

%% file: newcomm.tex
\newcommand{\be}{\begin{eqnarray}}
\newcommand{\ee}{\end{eqnarray}}
\def\lsim{\mathrel{\rlap{\lower4pt\hbox{\hskip 0.5 pt$\sim$}}
\raise1pt\hbox{$<$}}}                
\def\gsim{\mathrel{\rlap{\lower4pt\hbox{\hskip1pt$\sim$}}
\raise1pt\hbox{$>$}}}

\newcommand{\apr}{{A^\prime}}

\makeatletter
\newcommand\arraybslash{\let\\\@arraycr}
\makeatother
\newcommand{\thedate}{\today}

\def\gevc2{(GeV/c)$^2$}
\newcommand*{\jlab}{Jefferson Lab, Newport News, VA 23606, USA}
\newcommand*{\nhs}{University of New Hampshire, Durham NH 03824, USA}

\newcommand*{\frascati}{Istituto Nazionale di Fisica Nucleare, Laboratori Nazionali di Frascati, P.O. 13, 00044 Frascati, Italy}
\newcommand*{\genova}{Istituto Nazionale di Fisica Nucleare, Sezione di Genova e Dipartimento di Fisica dell'Universit\`a, 16146 Genova, Italy}

\newcommand*{\lasapienza}{Istituto Nazionale di Fisica Nucleare, Sezione di Roma e  Universit\`a La Sapienza, 00185 Roma, Italy}
\newcommand*{\infnct}{Istituto Nazionale di Fisica Nucleare, Sezione di Catania e Dipartimento di Fisica dell'Universit\`a, Catania, Italy}
\newcommand*{\infnbo}{Istituto Nazionale di Fisica Nucleare, Sezione di Bologna e Dipartimento di Fisica dell'Universit\`a, 40127 Bologna, Italy}
\newcommand*{\infnba}{Istituto Nazionale di Fisica Nucleare, Sezione di Bari e Dipartimento di Fisica dell'Universit\`a, Bari, Italy}
\newcommand*{\infnfe}{Istituto Nazionale di Fisica Nucleare, Sezione di Ferrara e Dipartimento di Fisica dell'Universit\`a, Ferrara, Italy}
\newcommand*{\infnle}{Istituto Nazionale di Fisica Nucleare, Sezione di Lecce 73100 Lecce, Italy}

\newcommand*{\ipn}{Institut de Physique Nucleaire d'Orsay, IN2P3, BP 1, 91406 Orsay, France}

\newcommand*{\ohio}{Ohio University, Department of Physics, Athens, OH 45701, USA}
\newcommand*{\gwu} {The George Washington University, Washington, D.C., 20052}

\newcommand*{\odu}{Old Dominion University, Department of Physics,
Norfolk VA 23529, USA}

\newcommand*{\edinb}{Edinburgh University, Edinburgh EH9 3JZ, United Kingdom}
\newcommand*{\glasgow}{University of Glasgow, Glasgow G12 8QQ, United Kingdom}

\newcommand{\sassari}{Universit\`a di Sassari e Istituto Nazionale di Fisica Nucleare, 07100 Sassari, Italy}
\newcommand{\torvergata} {Istituto Nazionale di Fisica Nucleare, Sezione di Roma-TorVergata e Dipartimento di Fisica dell'Universit\`a, Roma, Italy}

\newcommand{\torino} {Istituto Nazionale di Fisica Nucleare, Sezione di Torino,  Torino, Italy}
\newcommand{\padova} {Istituto Nazionale di Fisica Nucleare, Sezione di Padova,  Padova, Italy}
\newcommand*{\hu}{Department of Physics, Hampton University, Hampton VA 23668, USA }
\newcommand*{\canisius}{Canisius College, Buffalo NY 14208, USA}
\newcommand*{\occidental}{Occidental College, Los Angeles, California 90041, USA}
\newcommand*{\unm}{University of New Mexico, Albuquerque, New Mexico, NM}
\newcommand*{\mainz}{Institut fur Kernphysik, Johannes Gutenberg-Universitat Mainz, 55128 Mainz, Germany}
\newcommand*{\slac}{Stanford Linear Accelerator Center (SLAC), Menlo Park, CA 94025, US}
\newcommand*{\fnal}{Center for Particle Astrophysics, Fermi National Accelerator Laboratory, Batavia, IL 60510}
\newcommand*{\msu}{Mississippi State University, Mississippi State, MS 39762, USA}
\newcommand*{\stony}{C.N. Yang Inst. for Theoretical Physics, Stony Brook University, NY}
\newcommand*{\idaho}{Dept. of Physics, Idaho State University, Pocatello, ID 83201  USA}
\newcommand*{\spaolo}{Instituto de Fisica, Universidade de S\~ao Paulo, Brasil}
\newcommand*{\northw}{Northwestern University, Evanston, IL 60208, USA}
\newcommand*{\julich}{Nuclear Physics Institute and Juelich Center for Hadron Physics, Forschungszentrum Juelich, Germany}
\newcommand*{\mito}{Massachusetts Institute of Technology, Cambridge, MA 02139, USA}
\newcommand*{\bnl}{Brookhaven National Laboratory, Upton, NY 11973, USA}

%% file: BDX-PAC45-upd-intro.tex
\section{Executive summary}
\label{sec:intro}
This document is a summary of the effort by the BDX collaboration to address the concerns raised by JLAB-PAC44 regarding the proposal {\it PR12-16-001 Dark matter search in a Beam-Dump eXperiment (BDX) at Jefferson Lab}~\cite{bdx-proposal}.

\begin{figure}[t!] 
\center 
\includegraphics[width=12.5cm]{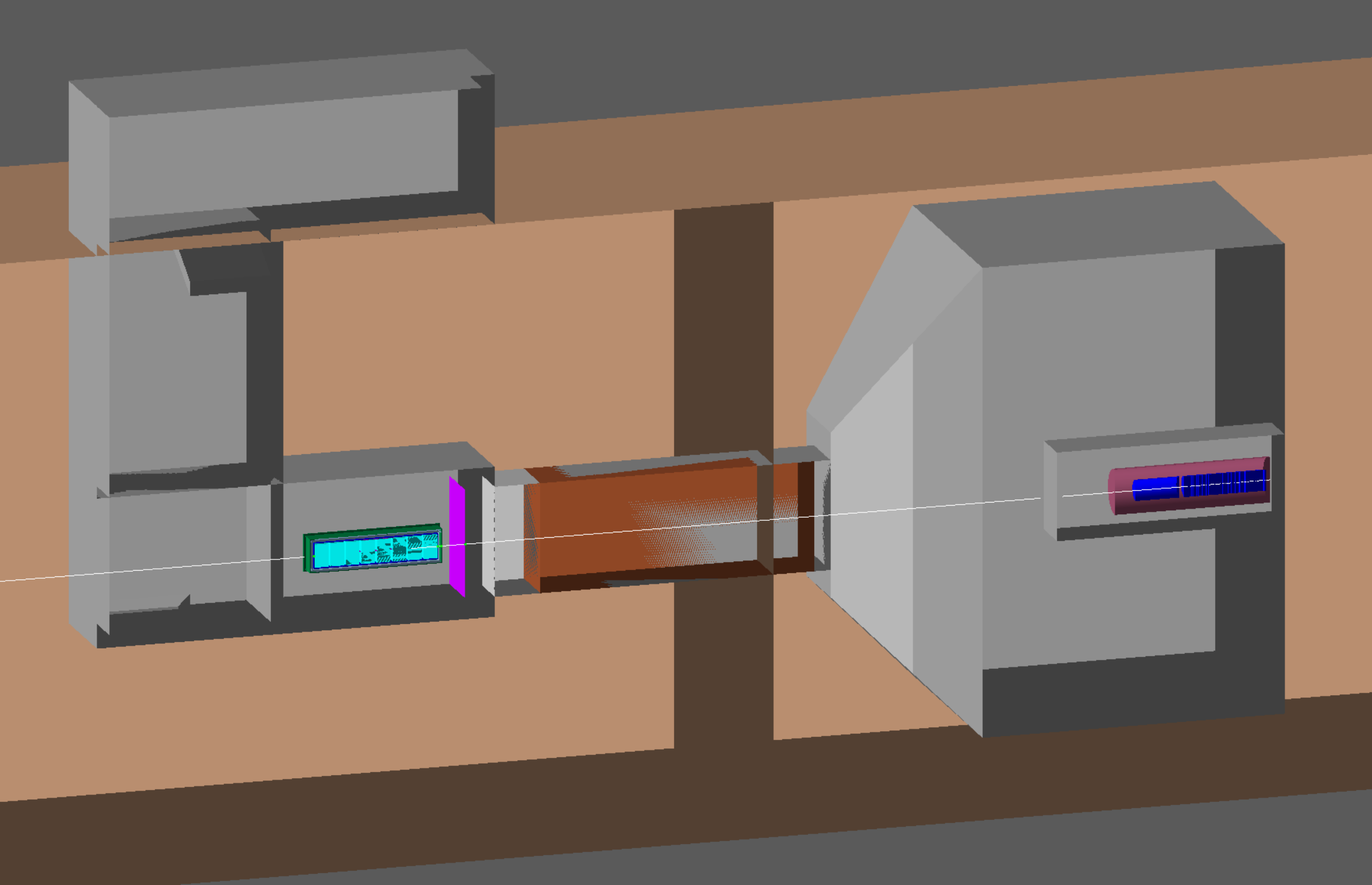}  
 \caption{The experimental set-up of BDX. The GEANT4 drawing shows (from the right) the beam-dump, the concrete bunker, the iron shielding, the new underground facility and the BDX detector. 
\label{fig:bdx_gemc} }
\end{figure}

We remind the reader that BDX aims to measure the scattering of (MeV - GeV) mass dark matter particles, $\chi$, produced in the interaction of the 11 GeV electron beam in the Hall A beam dump, off atomic electrons. The scattering is detected 
by recording the high energy (E$_{Dep}>$ 500 MeV) electromagnetic shower produced in the interaction.
The proposed experimental set-up includes a new underground facility to be built about 20 m downstream of the Hall-A beam-dump. The facility will host the BDX detector made of modules of CsI(Tl) crystals to detect the electromagnetic shower and two layers of active veto  to reject cosmic and beam-on background.  A passive shielding made of $\sim$6 m of iron will range-out almost all Standard Model particles produced in the beam/beam-dump interaction. Figure~\ref{fig:bdx_gemc} shows a sketch of the BDX set-up as implemented in GEANT4.

To address the concerns  we have improved our simulation tools and developed a specific plan to make test measurements of the muon flux downstream of the Hall A beam dump. The test measurements will be made in holes bored into the earth 
down to the beam height
at the location of the proposed facility. The determination of the muon flux is straightforward both experimentally and regarding its interpretation as originating from interactions in the electron beam dump. 
Both initiatives have benefited from advice and expertise from the JLab Radiation Control group. \

Simulations of the BDX experiment are now being conducted using the program FLUKA in addition to GEANT4, which was used to simulate interactions for the proposal. 
Included in the new FLUKA Monte Carlo simulations are the full geometry of passive and active materials, all physics processes and a tuned set of biasing weights to speed up the running time while preserving the results accuracy. This method can provide a statistical accuracy comparable to what is obtained generating  the electrons on target (EOT) collected by the experiment (N$_{EOT}\sim 10^{22}$). These studies were performed in collaboration with  JLab Radiological Control Group, recognized experts in background simulation and estimate. FLUKA results  were found to be in good agreement with lower statistics  GEANT4 simulations  
 and confirmed that neutrinos are the only source of expected beam-related background since the other particles are either ranged-out (muons, gamma and electrons) by the planned shielding  or  do not deposit enough energy into the BDX detector to trigger the DAQ (neutrons).  The results of these extended studies confirm the background expectations presented in the proposal. \

We also have developed a detailed plan to determine  the forward-going muon flux and the prompt  background in the current CEBAF + Hall-A dump configuration. Although it will not be possible to directly compare results of this test with the 
experimental set-up proposed in PR12-16-001 that will make use of increased shielding, the measurement will be extremely useful to validate the Monte Carlo simulation tools (GEANT4 and FLUKA) used to design the new underground facility 
and optimize the BDX detector.\
The muon  flux will be sampled at different beam heights at two distances from the beam dump covering a range of angles downstream of the beam-dump to map-out the radiation field to compare to simulations.
The fluxes will be measured with a detector package (BDX-Hodo) containing a CsI(Tl) crystal from the  BDX electromagnetic calorimeter
surrounded by plastic scintillator paddles to trigger on horizontal muons and veto cosmic rays. The BDX-Hodo is specifically designed for this measurement and will use a  loose trigger to provide further information on 
beam-related  low energy background  (dominated by neutrons). Characterization of  $\mu$ production in the beam-dump  will help in  understanding $\nu$ backgrounds, which are expected to be the irreducible background in our experiment. \

This document is organised as follow: a brief theoretical update as well as a brief summary of new results from other experiments since the 2016 PAC44 meeting is provided in Sec.~\ref{sec:th}; 
FLUKA beam-on background  for BDX experiment and comparison with GEANT4 results are reported in Sec.~\ref{sec:sim}.
The proposed beam-on background test measurements  in the existing experimental configuration are reported in Sec.~\ref{sec:mu}.

%% file: BDX-PAC45-upd-th.tex
\section{Theory update}
\label{sec:th}

\subsection{Review of Light Thermal Dark Matter}
\label{sec:theory}
In this section we review representative models of sub-GeV Dark Matter (DM) as presented more comprehensively in Refs.~\cite{bdx-proposal}.
If the dark and visible matter have sufficiently large interactions to achieve thermal equilibrium during the early universe, the resulting DM abundance greatly exceeds the observed density in the universe today; thus, a  thermal origin requires a sufficient DM annihilation rate to deplete this excess abundance and agree with observation at later times. For thermal dark matter below the GeV scale, this requirement can only be satisfied if the dark sector contains comparably light new force carriers to mediate the necessary annihilation process. Such ``mediators" must couple to visible matter and be neutral under the Standard Model (SM) gauge group. A popular representative model involves a dark photon $\apr$ with mass $m_{\apr}$ and Lagrangian
\cite{Holdom:1985ag}
 \be
 \label{eq:lagrangian}
{\cal L} =
-\frac{1}{4}F^\prime_{\mu\nu} F^{\prime\,\mu\nu} + \frac{\epsilon}{2} F^\prime_{\mu\nu} F_{\mu \nu} + \frac{m^2_{A^\prime}}{2} A^{\prime}_\mu A^{\prime\, \mu} + g_D \apr_\mu J^\mu_\chi  +  e A_\mu J^\mu_{\rm EM}   ,
\ee
where  $F^\prime_{\mu\nu} \equiv \partial_\mu A^\prime_\nu -  \partial_\nu A^\prime_\mu$ is the dark photon field strength,
$F_{\mu\nu} \equiv \partial_\mu A_\nu -  \partial_\nu A_\mu$ is the electromagnetic field strength,
  $g_D \equiv \sqrt{4\pi \alpha_D}$ is the dark gauge coupling, and $J^\mu_\chi$ and $J^\mu_{\rm EM}$ are the DM and electromagnetic matter currents, respectively. Here $\epsilon$ parametrizes the degree of kinetic mixing between dark and visible interactions. 
In this class of models, SM fermions acquire an effective ``milli-charge'' $\epsilon e$ under the short-range force carried by $\apr$.
 The phenomenology of the DM interaction depends on the DM/mediator mass hierarchy and on the details of the dark current $J^\mu_\chi$. If there 
 is only one dark sector state, the dark current generically contains elastic interactions with the dark photon. However, if there are two (or more) dark sector states the dark photon can couple to the dark sector states off-diagonally, as we will illustrate shortly. This latter scenario can lead to distinct signatures, which beam-dump experiments are especially suited for.

\subsubsection{Predictive Thermal Targets }
 In the paradigm of a thermal origin for DM, DM would have acquired its current abundance through annihilation directly/indirectly into the SM. Here, we focus on the direct annihilation regime, in which $m_{\chi} < m_{\rm MED.}$, where $m_{\rm MED.}$ would correspond to $m_\apr$ in the model we are focusing on. In this case, the thermal relic abundance is achieved via $\chi \bar \chi \to ff$, where $f$ are SM fermions. This annihilation rate scales as:
\be
\hspace{1cm}({\rm direct ~annihilation})~~~~\langle \sigma v (\chi \chi \to ff) \rangle   \propto     \epsilon^2  \alpha_D  \left( \frac{  m_{\chi}}{   m_{ A^\prime}  \!\!} \right)^4~~ ,~~~~~~~~~  ~~~~
\ee
and offers a  predictive target for discovery or falsifiability since the dark coupling $\alpha_{D}$  and mass ratio $m_{{\chi}}/m_{\apr}$ are 
at most ${\cal O}(1)$ in this $m_{\apr} > m_{\chi}$ regime, so there is a {\it minimum} SM-mediator coupling compatible with a thermal history; larger values of $g_D$ require non-perturbative dynamics in the mediator-SM coupling or intricate model building.

\subsubsection{Important Variations}
We now consider two important variations: in the dark sector matter states, and in the mediator nature, respectively. First, we consider a possibility that can arise in the representative model from Eq.~\ref{eq:lagrangian} {\it without} any additional particle content under the general assumption that the dark sector matter is sensitive to dark symmetry breaking. Second, we consider a variation on the nature of the mediator where instead of kinetically mixing with hypercharge, the mediator is the gauge field of a theory in which one gauges lepton number.

{\bf Inelastic Dark Matter (iDM)}

If the $A^\prime$ couples to a DM fermion with both Dirac and Majorana masses, the leading interaction is generically off-diagonal and 
\be
A_\mu^\prime  J^\mu_{DM} ~ \to ~ A_\mu^\prime  \bar \chi_1 \gamma^\mu \chi_2~,~
\ee
where the usual Dirac fermion $\chi$ decomposes into two Majorana (``pseudo-Dirac") states $\chi_{1,2}$ with masses $m_{1,2}$ split
by an amount $\Delta$.This kind of scenario is well motivated for LDM, as it can arise in the theory in the general case when the dark sector matter is sensitive to the dark sector spontaneous symmetry breaking. Moreover, inelastic dark mater models are safe from CMB constraints \cite{Izaguirre:2015yja}, and have striking implications for possible signatures at BDX. \\

{\bf Leptophilic $A^\prime$ and Dark Matter}

A similar scenario involving a vector mediator arises from gauging the difference  between electron and muon numbers under the abelian $U(1)_{e-\mu}$ group ---  by gauging the difference in two lepton numbers we ensure that the theory is anomaly free.
Instead of kinetic mixing, the light vector particle here has direct couplings to SM leptonic currents 
\be
A_\beta^\prime J^\beta_{SM} ~\to~   g_V   A_\mu^\prime \left(  \bar e \gamma^\beta e  +  \bar \nu_e \gamma^\beta \nu_e -
\bar \mu \gamma^\beta \mu  +  \bar \nu_\mu \gamma^\beta \nu_\mu \right)  ~,~~
\ee
where $g_V$  is the gauge coupling of this model, which we normalize to the electric charge,  $g_V \equiv \epsilon e$ and consider
parameter space in terms of $\epsilon$, like in the case of kinetic mixing. Note that here, the $A^\prime$ does not couple
to SM quarks at tree level, but it does couple to neutrinos, which carry electron or muon numbers.
Note also that this scenario is one of the few combinations of SM quantum numbers that can be gauged without requiring additional field content.
Assigning the DM $e-\mu$ number yields the familiar $g_D A^\prime_\beta J^{\beta}_{\rm DM}$ interaction as in Eq.~\ref{eq:lagrangian}.
 Both of these variations can give rise to thermal  Light Dark Matter (LDM) as discussed above.

\subsection{New Experimental Results}
Since the original BDX proposal was submitted to the JLab PAC in July of 2016, the experimental landscape has changed somewhat 
due to new results in the intervening time period. These new results collectively rule out the $g-2_\mu$ explanation by a kinetically-mixed dark photon that decays into fully invisible final states. We summarize these below, but we note that BDX remains an extremely compelling experiment because it is uniquely suited to probe new parameter space that's compatible with a thermal origin as well as the structure of the Dark Sector, and importantly this can be done in the very near future.

\subsubsection{NA64}
The NA64 collaboration has recently reported a new search for invisibly decaying dark photons produced via bremsstrahlung in a secondary 
electron beam at the CERN SPS \cite{Banerjee:2016tad}. The detector consists of a magnetic spectrometer (tracker with associated bending magnet), followed by a calorimeter system composed 
of an ECAL, a veto, and a highly hermetic HCAL. The dark mediator is directly produced in the ECAL, and the signal region is defined as a reconstructed track, an energy deposition in the ECAL below a roughly half the original beam energy, and no activity in the veto or the HCAL. 

\subsubsection{MiniBooNE}
The MiniBooNE experiment has reported new limits on light dark matter produced via $\pi^0 / \eta \to \gamma \apr \to \gamma \chi \chi$ decays \cite{Aguilar-Arevalo:2017mqx}. The 8 GeV proton beam was run in off-target mode, whereby the beam was steered to impinge on a steel beam dump. This strategy was pursued to reduce the neutrino flux by an order of magnitude. MiniBooNE is a 800 ton mineral oil Cherenkov detector situated 490 m downstream of the beam dump. The experimental 
sensitivity is limited by the uncertainties on the neutrino background. First results based on $1.8\times10^{20}$ POT have been published for DM-nucleon scattering. 

\begin{figure}[t!]
\center
\includegraphics[width=8.2cm]{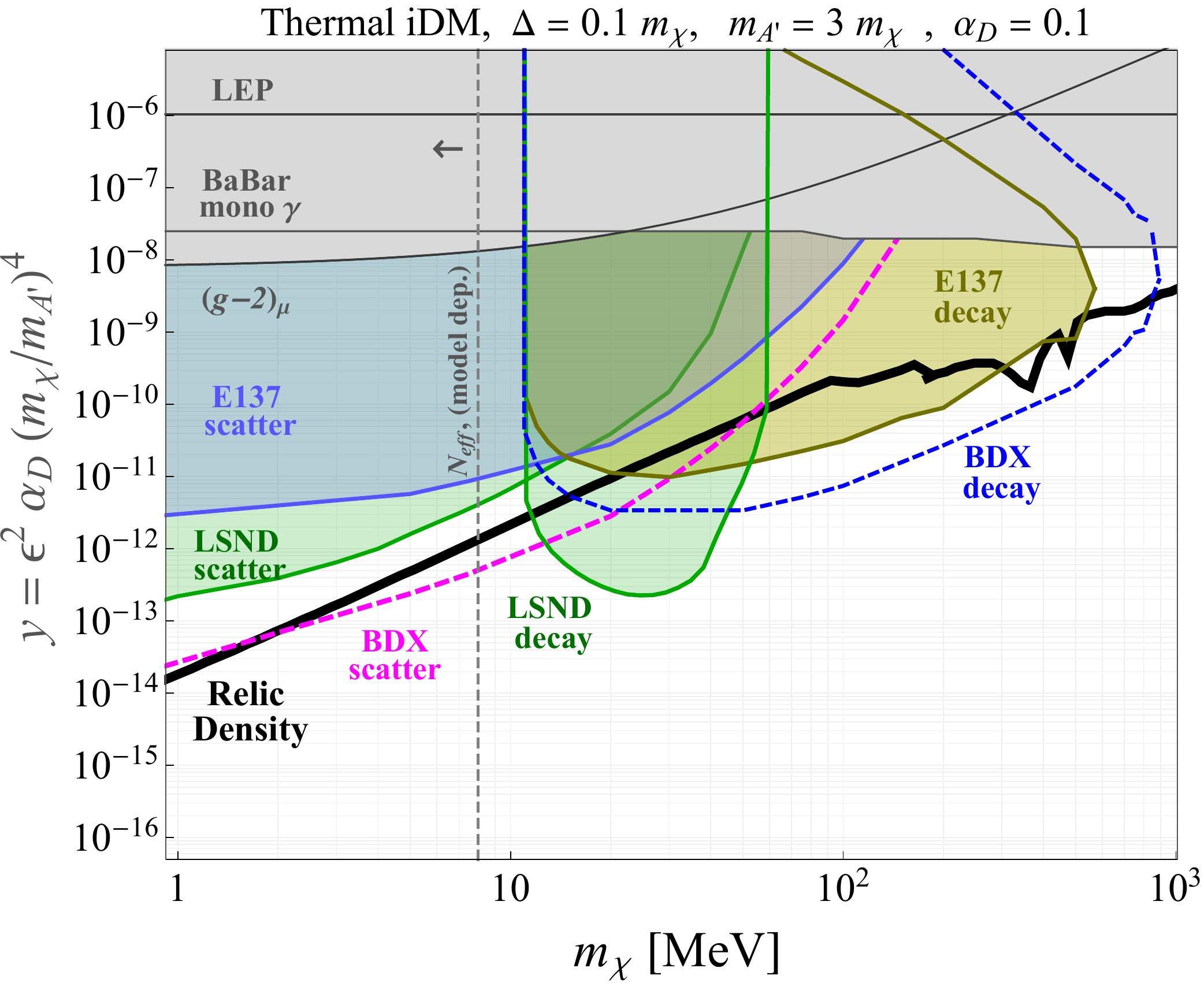}  
\includegraphics[width=8.2cm]{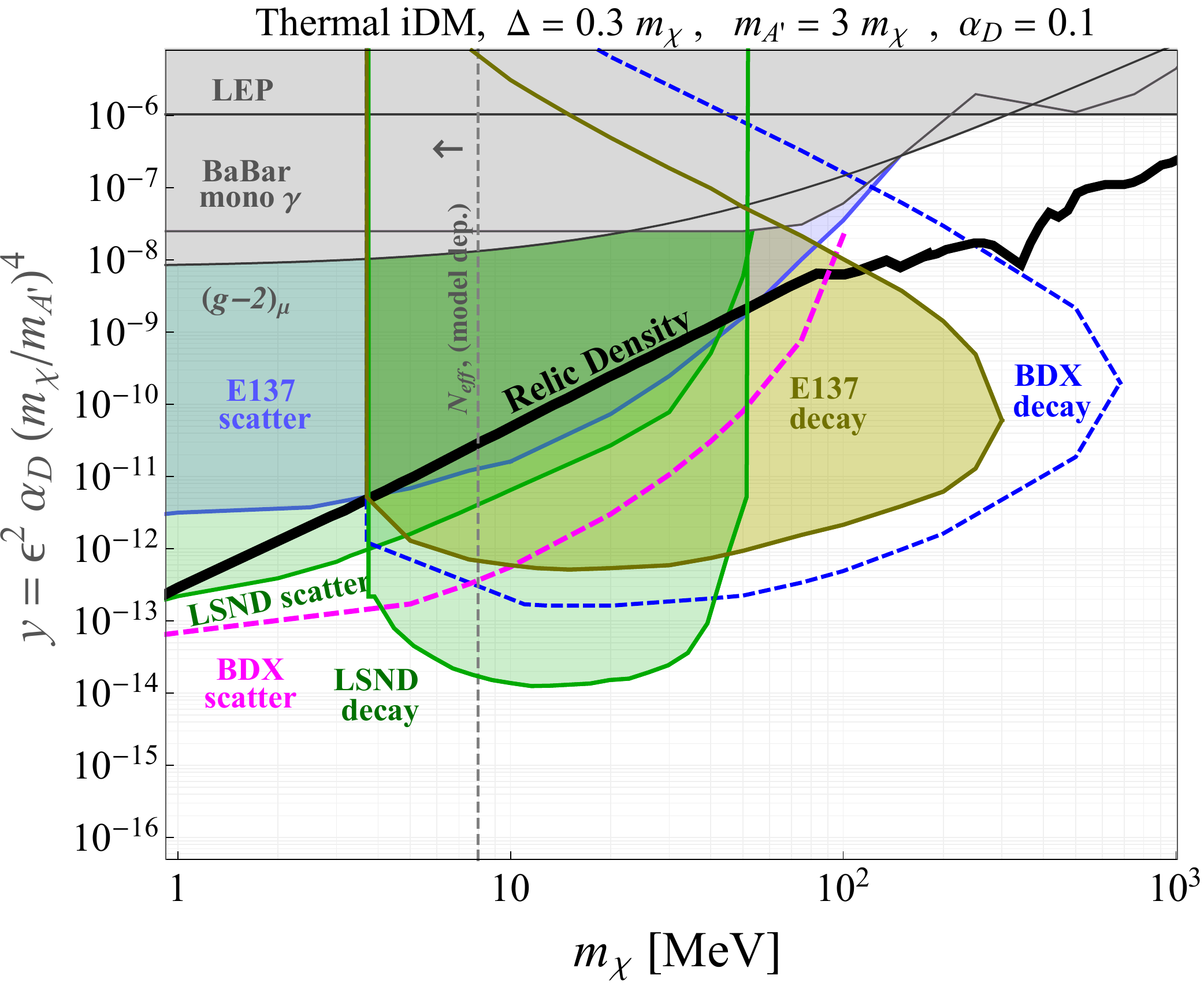} 
\caption{ Plot of BDX yield projections for inelastic DM scattering $\chi e^- \to \chi e^-$ (pink dashed) and decay $\chi_2 \to \chi_1 e^+e^-$ signatures for  $10^{22}$ electrons on target. Two different mass splittings are shown: 10\% (left) and 30\% (right). The blue dashed line shows BDX reach. For details see \cite{Izaguirre:2017bqb} .  }
\label{fig:idm}
\end{figure}
\subsubsection{BaBar}
The latest result from the BaBar collaboration \cite{Lees:2017lec} presents new limits on invisibly decaying dark photons via 
radiative return $e^+e^- \to \gamma \apr \to \gamma (\apr \to \chi \chi)$ at SLAC. The measurement performed consists of a bump hunt in the missing mass spectrum. The results of the experiment rule out $\epsilon < 10^{-3}$ across the MeV-10 GeV $m_{A'}$ regime.

\subsection{New Inelastic DM Phenomenology at BDX }
In addition to new experimental results, new work on the theory side has fully fledged out the phenomenology of models of inelastic dark matter at fixed target facilities, and highlighted the unique sensitivity that beam-dump experiments can have to these scenarios. Ref.~\cite{Izaguirre:2017bqb} proposed a series of searches that can exploit these features. In particular, proposed experiments like BDX can be sensitive to the de-excitation of long-lived excited states produced in the beam dump. Here, in the aforementioned model of inelastic dark matter, the excited state $\chi_2$ de-excites via $\chi_2 \rightarrow \chi_1 \ell^+ \ell^-$. In fact, for $\alpha_D$ and $\epsilon$ values that are consistent with thermal parameter space, the $\chi_2$ lifetime is macroscopic and such decays could occur inside the BDX detector at an appreciable rate. Fig.~\ref{fig:idm} illustrates the possible sensitivity in a background-limited signal region that requires the exited state to deposit $E_e > 300$ MeV inside BDX.

\subsection{Accelerated Broader Interest in the Community}

The last year has also seen considerable interest in the broader community to launch a vibrant  experimental program to test light dark matter. This is evidenced by the 100$^+$ attendees at the Dark Sectors 2016 workshop last summer, culminating in a status report of this rapidly growing field \cite{Alexander:2016aln}. Moreover, in accordance with the P5 (Particle Physics Project Prioritization Panel) recommendation to devote resources to funding small experiments, the DOE is interested in identifying new, small projects for dark matter searches in areas of parameter space not currently being explored. DOE requested a community-organized workshop ``U.S. Cosmic Visions: New Ideas in Dark Matter'', which was held in early 2017 at the University of Maryland, to examine the next experimental steps in the search for dark matter. Finally, this community effort will result in a White Paper summarizing the science priorities. New small-scale accelerator experiments like BDX can play a crucial role in achieving the science goals discussed at the recent community-organized workshop.

\subsection{Summary of theoretical update}

In summary, BDX is uniquely suited to make timely progress in the quest for sub-GeV thermal-origin DM. It is the ultimate electron beam-dump experiment, whose sensitivity will only be limited by the beam-originate irreducible neutrino floor. In fact, BDX will be the first electron fixed-target experiment that reaches the neutrino irreducible floor. BDX will be able to test new parameter space consistent with a thermal origin, will be sensitive to the DM-mediator coupling, and in the event of a discovery, it could even start doing Dark Sector spectroscopy --- by measuring the lifetime of unstable states in the Dark Sector.

%% file: BDX-PAC45-upd-sim.tex
\section{FLUKA simulations of the BDX experiment}
\label{sec:sim}
In this Section we report the results obtained with FLUKA for the beam-on background expected in the BDX experimental configuration and its comparison to GEANT4 simulations. We show that FLUKA and GEANT4 agree quite well  confirming  results reported in Sec.~4.3 of PR12-16-001:  neutrinos are the only source of beam-related background since the other particles are either ranged-out (muons, gamma and electrons) by the planned shielding  or  do not deposit sufficient energy into the BDX detector to trigger the DAQ (neutrons).

\subsection{FLUKA results}  
\begin{figure}[ht!] 
\center
\includegraphics[width=14.5cm]{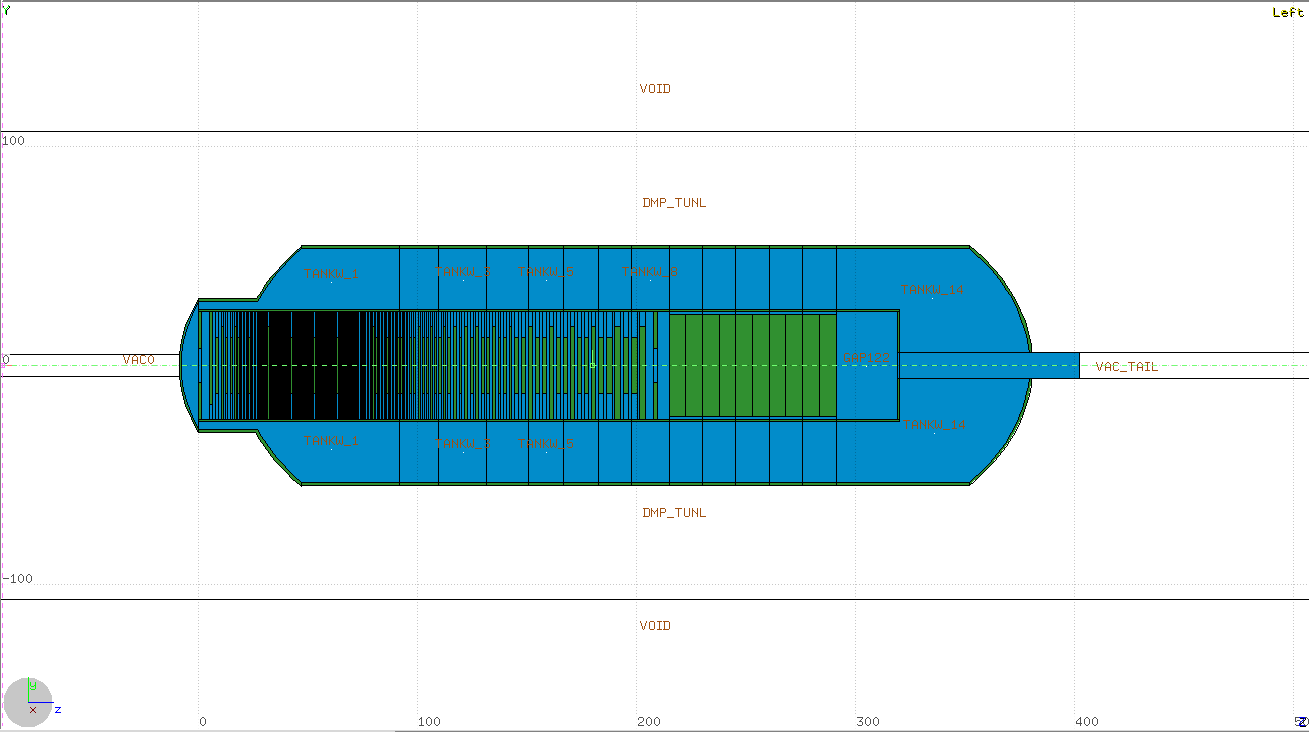}
\caption{Hall-A beam-dump implementation in FLUKA.}
\label{fig:fluka-bd}
\end{figure}

Starting from the current configuration of  the Hall-A beam dump geometry and materials
 implemented in FLUKA-2011.2c.5 by the Jefferson Lab Radiation Control Department~\cite{jnote-bd}, we added the iron shielding and the other components of the  BDX facility.  The input card used to run the program  includes all physics processes and a tuned set of biasing weights to speed up the running time while preserving the  results accuracy. 
We simulated an 11 GeV electron-beam interacting with the beam-dump, propagated all particles to the location of interest sampling the flux in different locations and compared to what was obtained by GEANT4.
The $\mu$, neutron, electron, and $\gamma$ fluence (differential in angle and energy) per electrons-on-target (EOT) were  calculated:
\begin{itemize}
\item{ at the exit of the dump (0.41 m  downstream of the beam window, through a circular area of 105 cm$^2$); }
\item{ in the  iron shielding (as an example we will show some energy spectra sampled in a 50x50 cm$^2$ flux detector located in the first half, 13.5 m  downstream of the dump entrance);}
\item{ at the front face location of the BDX detector (20.8m downstream of the beam-dump entrance).}

\end{itemize}
These points match the positions obtained with GEANT4 and shown in Fig.~26 of PR12-16-001. 
Figure~\ref{fig:fluka-bd} shows the FLUKA graphic representation of the BDX set-up implementation in FLUKA.

\subsubsection{Background in the beam-dump vault}
In our model we extended the original JLab Rad Con beam-dump description, by including  
 a more detailed  geometry and material composition around and downstream of the beam-dump.
\begin{figure}[ht!] 
\center
\includegraphics[width=12cm]{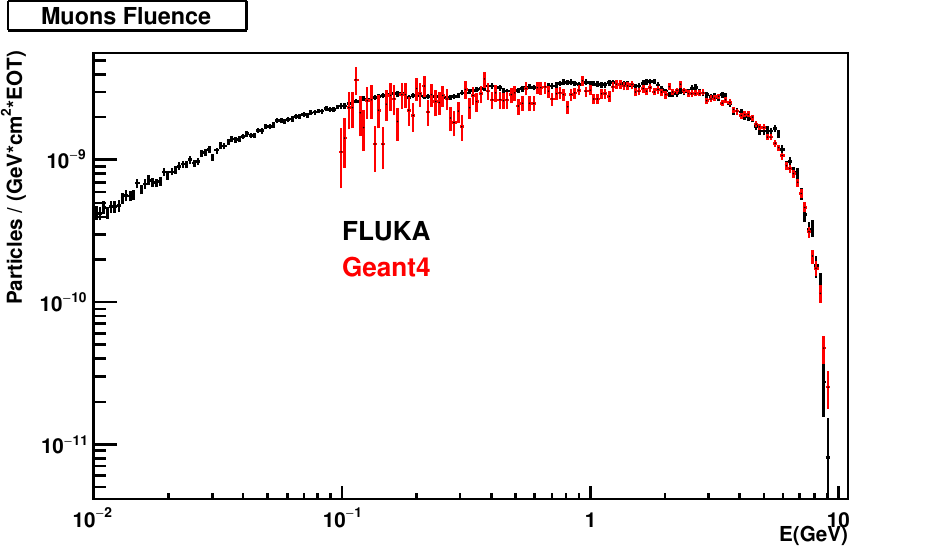}
\caption{Muon fluence at the exit  of the beam-dump obtained by FLUKA (black) and GEANT4 (red). The GEANT4 simulations have an energy cut  of  E$\mu=$100 MeV.} 
\label{fig:mu-flu-bd}
\end{figure}

\begin{figure}[ht!] 
\center
\includegraphics[width=7cm]{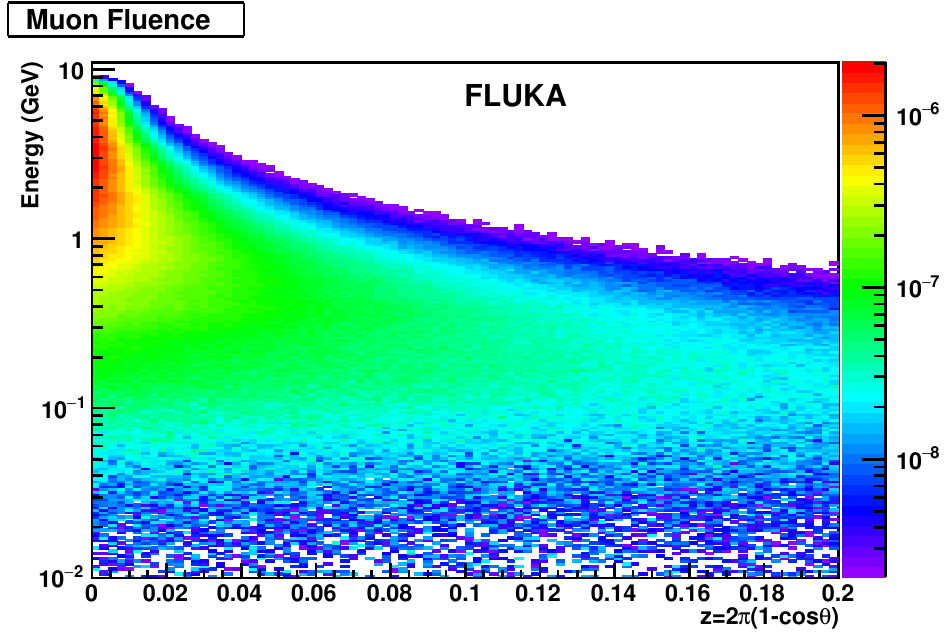} 
\includegraphics[width=7cm]{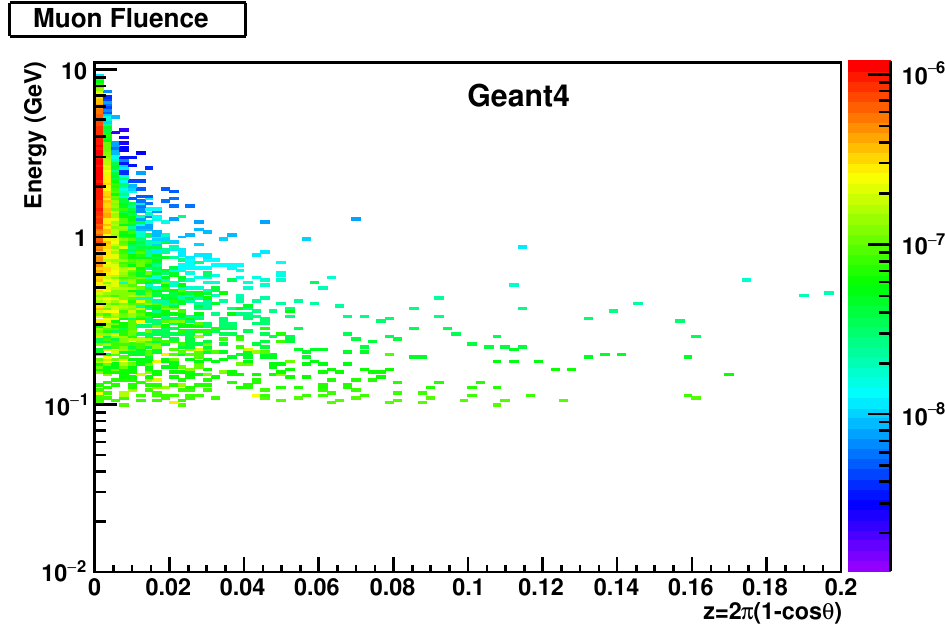} 
\caption{Energy vs azimuthal angle of muons crossing the flux detector located at the exit of the beam-dump obtained by FLUKA (left)  and GEANT4 (right). The distributions are compatible within the statistical uncertainty.}
\label{fig:mu-flu-bd-2d}
\end{figure}

\begin{figure}[ht!] 
\center
\includegraphics[width=8.cm]{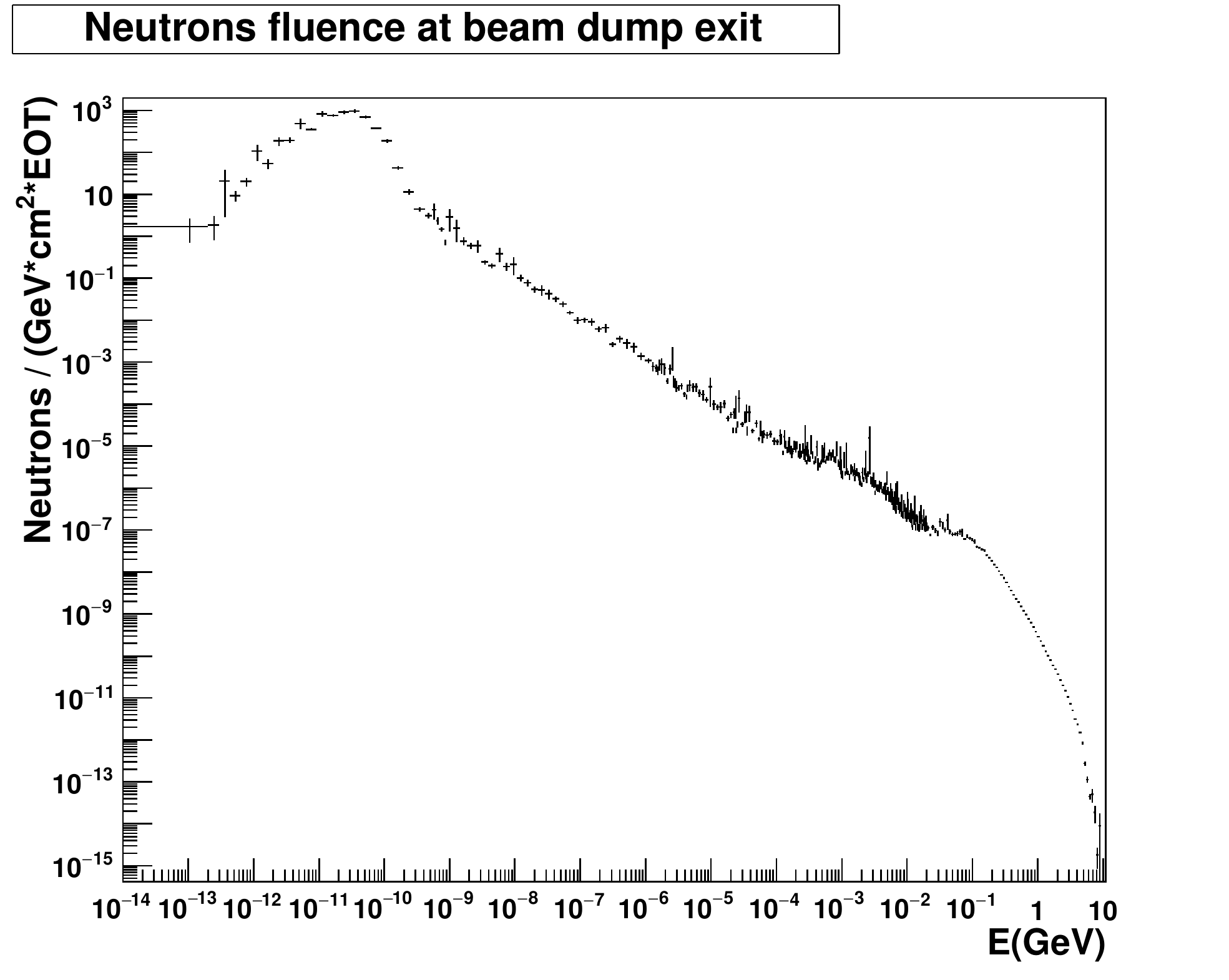}    
\includegraphics[width=8.cm]{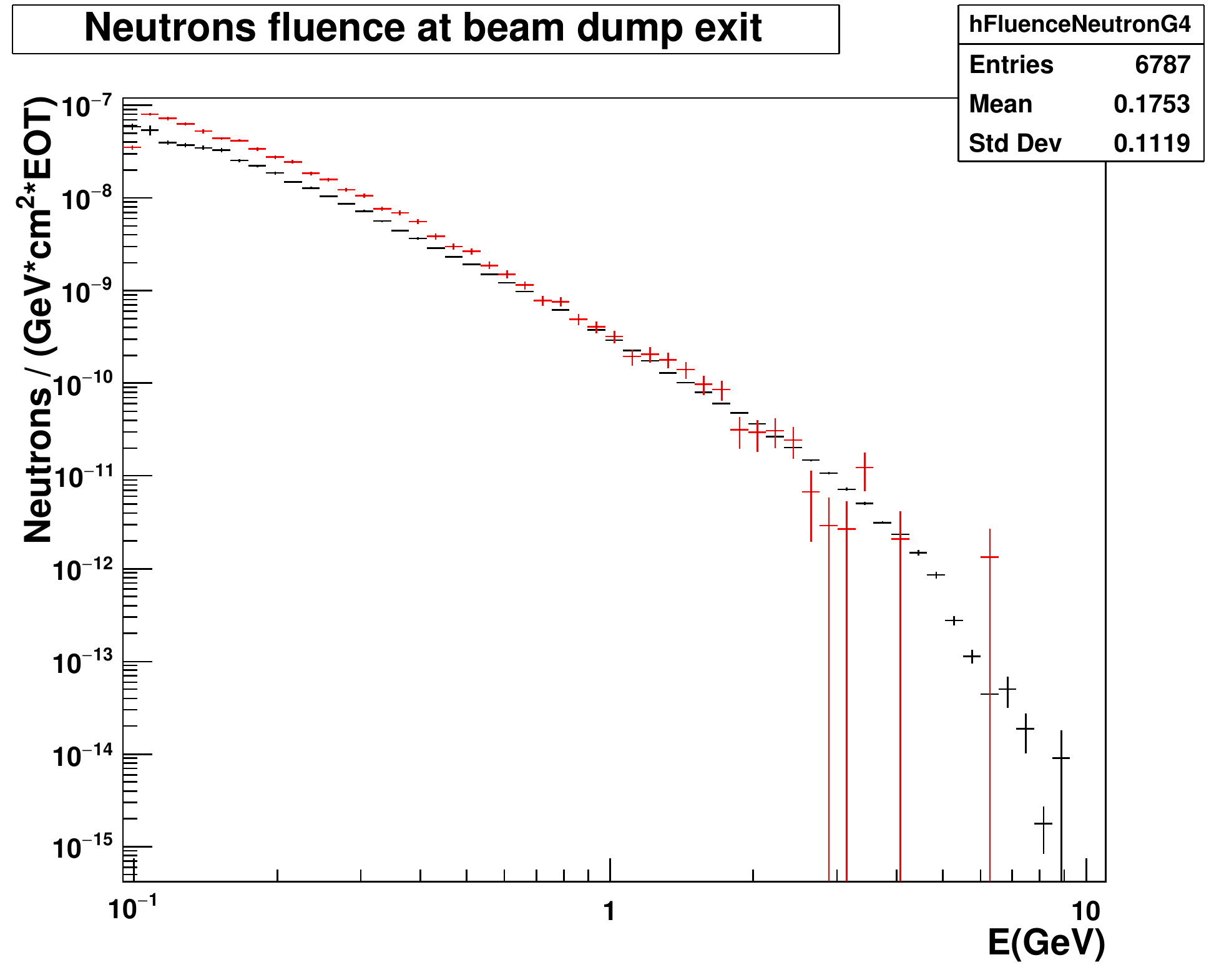} 
\caption{Neutron spectrum at the exit of the the dump obtained by FLUKA simulation (left). The right panel shows  the comparison between FLUKA (black) and GEANT4 (red) for the high energy part of the spectrum.}
\label{fig:n-comp}
\end{figure}

A comparison of muon fluence at the exit  of the beam-dump
obtained by FLUKA and GEANT4 are reported in Fig.~\ref{fig:mu-flu-bd}. Considering that low energy muons do not exit from the concrete beam-dump vault, to keep the GEANT4 running time reasonable, only particles with energy greater than 100 MeV have been tracked and sampled. A total of 1.3$\times10^{10}$ ($9\times 10^6$) EOT have been simulated with GEANT4 (FLUKA). The comparison of the two simulations shows a good agreement in the full energy range where data were generated.
In spite of a factor of $\times$100  less statistics, FLUKA shows, as expected, smaller error bars. This reflects the optimised biasing weights used by the simulation to generate high statistics for the low probability processes keeping the total number of events limited.
Figure~\ref{fig:mu-flu-bd-2d} shows the correlation between the muon energy and the azimuthal angle (with-respect-to the beam axes): the regions  populated in both simulations show a similar shape.
\begin{figure}[ht!] 
\center
\includegraphics[width=14.5cm]{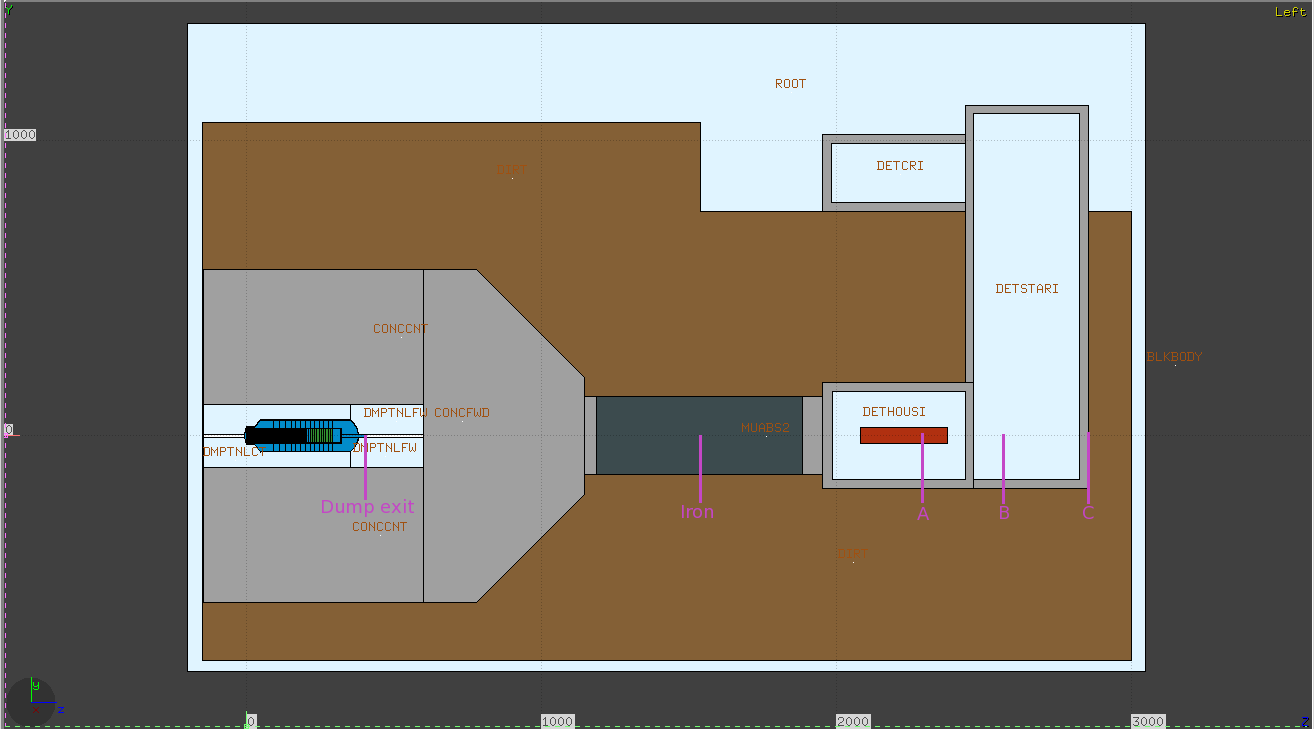} 
\caption{BDX underground facility in FLUKA.
\label{fig:fluka-underg}
}
\end{figure}

The neutron spectrum obtained with FLUKA sampled in the same position  is shown in the left panel of Fig.~\ref{fig:n-comp}. The right panel of the same figure shows the   comparison with GEANT4 for the high energy part, T$_n>$ 100 MeV. The reasonable  agreement (within a factor of con) indicates  that, in this energy range, both simulation tools are reliable. 

It is worth noting that background fluxes reported in this paragraph (in particular muons) can be tested in a dedicated measurement with the current JLab configuration.

\subsubsection{Background in the BDX underground facility }
\begin{figure}[ht!] 
\center
\includegraphics[width=7cm]{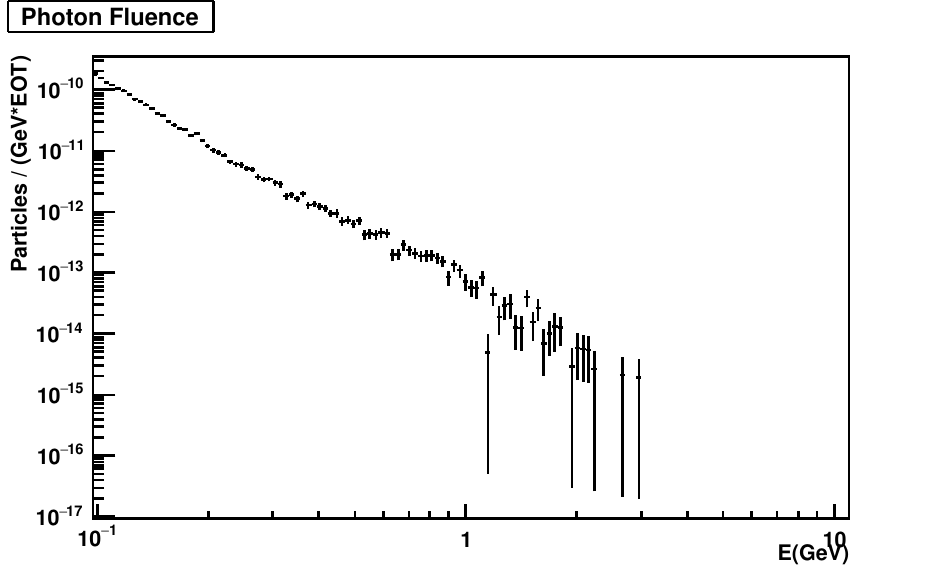}
\includegraphics[width=7cm]{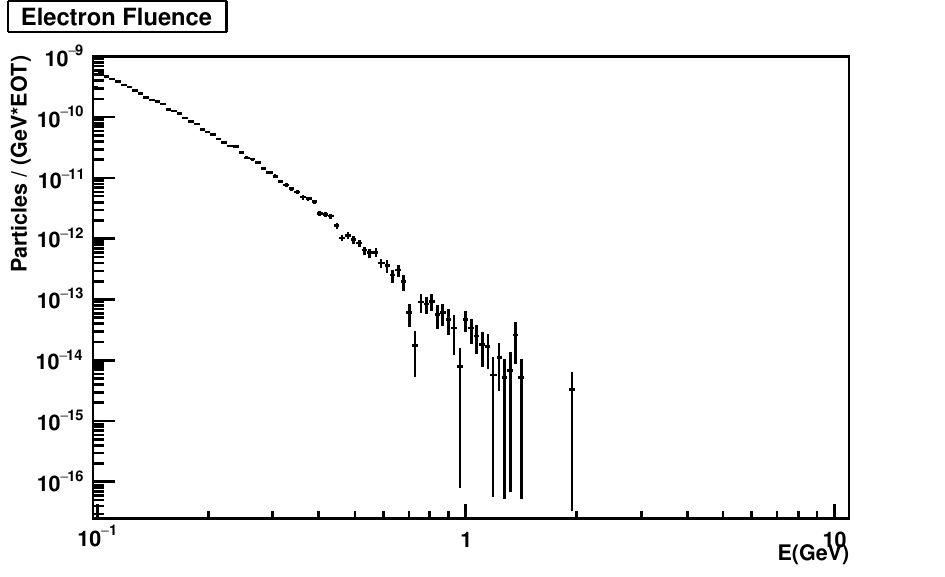}
\caption{Energy spectra of $\gamma$ (left) and electrons (right) sampled in the iron at 13.5 m from the beam-dump entrance as obtained from FLUKA. Due to the limited statistics, only FLUKA results are available.
\label{fig:gamma-elect-spec}}
\end{figure}

Figure~\ref{fig:fluka-underg} shows the implementation of the BDX experimental underground facility in FLUKA. Magenta lines indicate the different positions where the background flux has been sampled.
In this simulation we generate a total of 4$\times10^9$  and 4.$6\times 10^8$ 11 GeV electron/beam-dump-interactions with GEANT4 and FLUKA, respectively.
Fluxes of muons ($\mu^+$ and $\mu^-$), electrons, photons, and neutrons with energy between 0.5 GeV and 11 GeV have been sampled in the location described above.
Where possible,  the comparison between GEANT4 and FLUKA is shown.
Figure~\ref{fig:gamma-elect-spec} shows the energy spectrum for gammas and electrons  sampled in the first half of the iron shielding (13.5 m downstream of the beam-dump entrance). Due to the limited  statistics, only FLUKA results are available.

\subsubsection{Neutrino background in the BDX underground facility }
\begin{figure}[ht!] 
\center
\includegraphics[width=14.5cm]{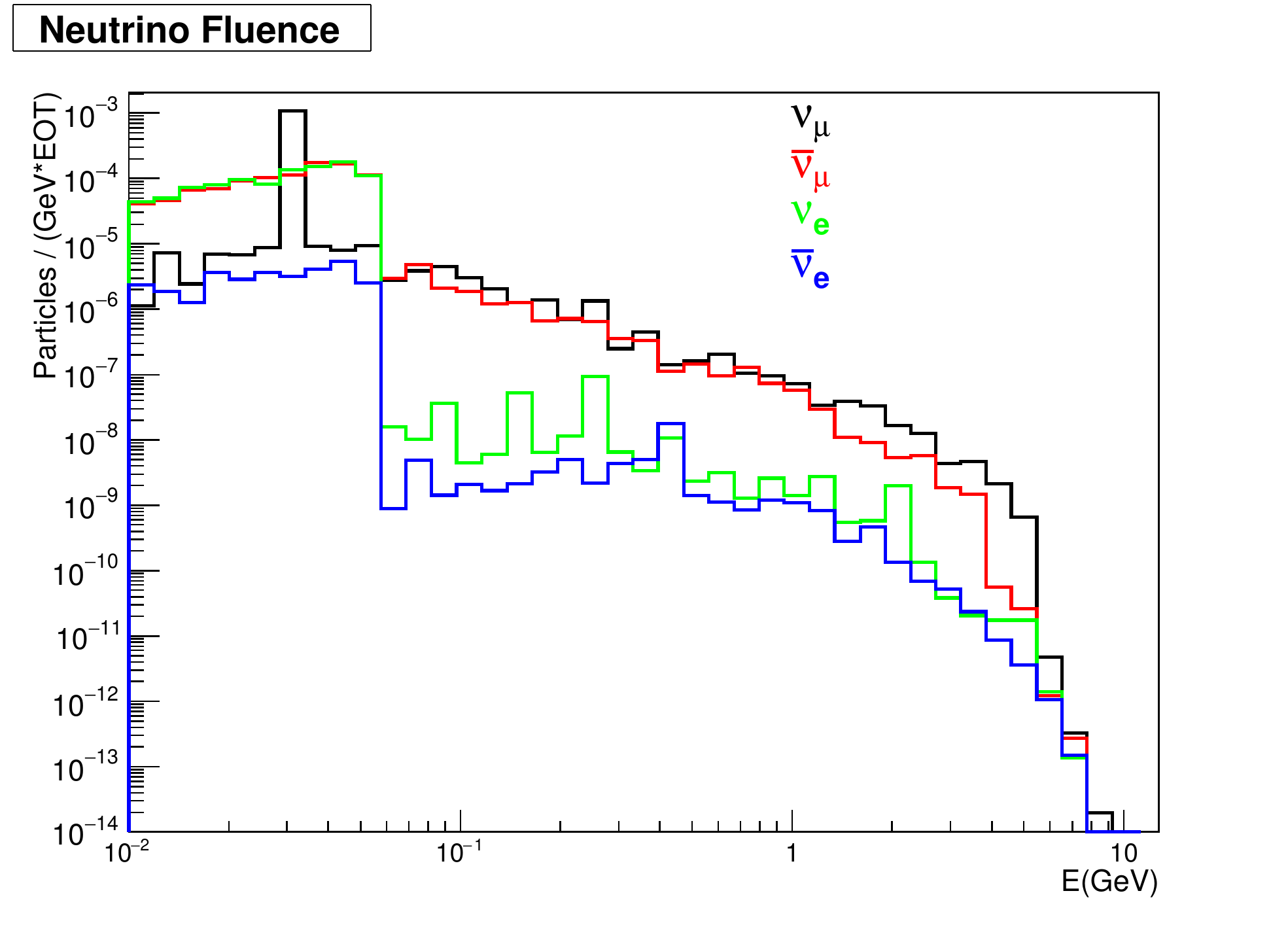} 
\caption{Energy spectrum of neutrinos ($\nu_{\mu}$, $\bar\nu_{\mu}$, $\nu_{e}$,  and $\bar\nu_{e}$)  impinging on the  BDX detector volume.}
\label{fig:neutflux-det}
\end{figure}

The four neutrinos species ($\nu_{\mu}$, $\bar\nu_{\mu}$, $\nu_{e}$,  and $\bar\nu_{e}$) were also tracked and sampled with FLUKA.  A sizeable number of neutrinos propagate to the BDX detector.  Figure.~\ref{fig:neutflux-det} shows the energy spectra  sampled at the detector front face.
 A tiny but not negligible part of the spectrum has energy greater than 500 MeV.
These events may produce a signal in the BDX detector similar to a DM interaction. 


Neutrinos ($\nu_{e}$, $\nu_{\bar e}$,  $\nu_{\mu}$, and   $\nu_{\bar \mu}$) are produced in muon decays and hadronic showers (pion decay). The majority comes from pion and muon decay at rest but a non negligible fraction, due to  in-flight pion decay, experience   a significant boost to  several GeV energy.
High energy neutrinos interacting with BDX detector by elastic and inelastic scattering may result in a significant energy deposition ($>$ 0.5 GeV) that may mimic an EM shower produced by the $\chi$-atomic electron interaction.
To estimate the neutrino background we used different simulation tools.   FLUKA has been used to generate neutrinos in 11 GeV electron/dump interaction, propagate them to the BDX location and sample particles  produced in the interaction with the active BDX volume (CsI(Tl)). Reaction products were then fed to GEANT4 code  that contains a detailed and realistic description of the BDX detector response.
Electron neutrinos and anti-neutrinos are suppressed by a factor of hundred with respect to $\nu_{\mu}$ and $\bar\nu_{\mu}$. These  (within a factor of 2-3) consistent to what was obtained using GEANT4 and presented in the proposal.
FLUKA only considers $\nu$-N and $\bar\nu$-N interactions disregarding  $\nu$-electron and $\bar\nu$-electron interaction.
We checked the validity of this approximation doing an analytical estimate of the $\nu$-e and $\bar\nu$-e contribution finding that for 10$^{22}$ EOT we are expecting less than 1 interaction on the BDX volume.

The implication  for the $\chi$-electron signal measurement for different $\nu$-matter interactions 
 are listed below.
\begin{itemize}
\item{$\nu_\mu$ N $\to \mu $ X: the Charge Current (CC) interaction produces a $\mu$ in the final state (beside the hadronic state X). This reaction can be identified and used to provide an experimental assessment of the $\nu_\mu$ background (and therefore estimate the $\nu_e$ contribution) by detecting a $\mu$ scattering in the detector (a MIP signal inside the calorimeter with or w/o activity in IV and OV) or, alternatively, selecting kinematics in which the $\mu$ is emitted at large angles.}
\item{$\nu_\mu$ N $\to \nu_\mu $ X: the Neutral Current (NC) interaction produces an hadronic state X that may interact in the detector (while the scattered $\nu$ escapes from detection). This can mimic an EM shower if $\pi^0$ ($\gamma$'s) are produced. However, due to the difference in mass, the scattered $\nu$ carries most  of the available energy providing a small transfer to the hadronic system and reducing the probability of an over-threshold energy deposition.}
\item{$\nu_e$ N $\to \nu_e $ X: same considerations as above.}
\item{$\nu_e$ N $\to e $ X: the CC interaction could produce a high energy electron into the detector that mimics the signal.  This background can be rejected considering again the different kinematics of the $\nu$ interaction with respect to the  $\chi$-electron scattering.  The significant difference in the polar angle (wrt the beam direction) allow to define a selection  cut to identify $\nu_e$ and separate from the $\chi$.
}
\end{itemize}

For a simulated statistics of  2.2$\times10^8$ EOT we obtained, after all rejection cuts and extrapolation to 10$^{22}$ EOT,  a background of 10 neutrino (3 $\nu_{\mu}$ + 0.2 $\bar\nu_{\mu}$ + 6$\nu_{e}$ + 0.7 $\bar\nu_{e}$) . The inventory and the conclusions are in good agreement with estimates obtained with  GEANT4 simulation reported in PR12-16-001
The rejection cuts optimization is in progress. Massive FLUKA simulations will be run in the next few months.
\begin{figure}[ht!] 
\center
\includegraphics[width=12.5cm]{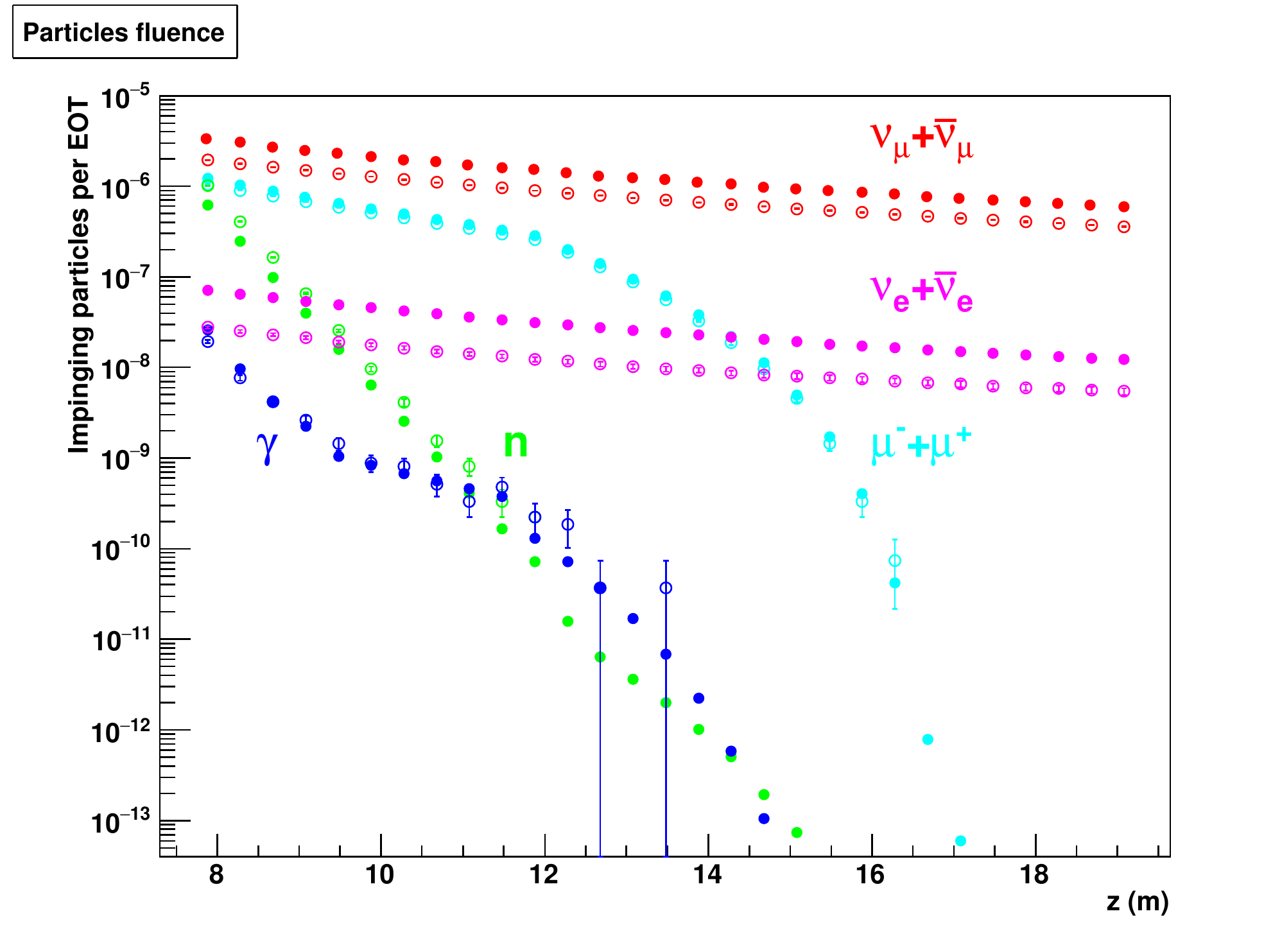}
\caption{Comparison of particles fluxes per EOT at different distances from the beam-dump entrance.  The position of the BDX detector is outside the graph range, at 20.6 m. Closed symbols refer to FLUKA results, while open symbols correspond to GEANT4 results.
\label{fig:summaryflux}}
\end{figure}

\subsection{Summary of simulations for BDX}
In this Section we reported results obtained with FLUKA to accurately simulate the interaction of the CEBAF 11 GeV electron-beam with the Hall-A dump and estimate the beam-on background in different locations in the fully shielded 
BDX experimental configuration.
Figure~\ref{fig:summaryflux} summarizes the results of this study: for energy larger than 500 MeV, only neutrinos produced in the electron-beam-dump interaction  propagate trough the proposed  shielding and reach the BDX detector.
Results were found to be consistent, within a factor of 2-3, with previous GEANT4 simulations confirming the findings  reported in PR12-16-001. For some locations, results were compared with JLab Radiological Control Group estimates finding a good agreement.  Making use of biasing weights FLUKA reduces the running time while preserving accuracy. We are planning to run massive FLUKA simulations in the next few months to generate a number of interactions 
similar to that expected in the experiment live time. 

In the next Section we propose a measurement behind the current Hall A beam dump, but in its present shielding configuration, to validate some of these findings.

%% file: BDX-PAC45-upd-mu.tex
\section{Test measurement of beam-on backgrounds}
\label{sec:mu}
A complete BDX  beam-on background assessment will only be possible when the new underground facility (including the planned iron shielding between the dump and the BDX detector) is built. Nevertheless, 
following  the PAC-44 recommendation, we propose to  measure the muon fluxes produced in the beam dump in the current experimental set up.
 FLUKA simulations have been used to estimate muon rates that can be measured in test pipes located behind Hall-A at beam level. Measuring the muon rates at this location will provide an absolute normalization of Monte Carlo simulations as well as confirm expected background rates for the present Hall-A beam dump geometry. Characterization of $\mu$ background  will provide some useful constraints on $\nu$  produced in the beam-dump. This test will be also validate the use of CsI(Tl) crystals read out by SiPM's and flash ADCs  in the noisy low-energy neutron environment of the BDX experiment.

In this Section we describe the  proposed measurement and expected results from this test.
All details, together with costs estimate and a detailed work plan, are reported in a dedicated Note~\cite{mutest-note}. 
\begin{figure}[tph] 
\center  
\includegraphics[width=11.5cm]{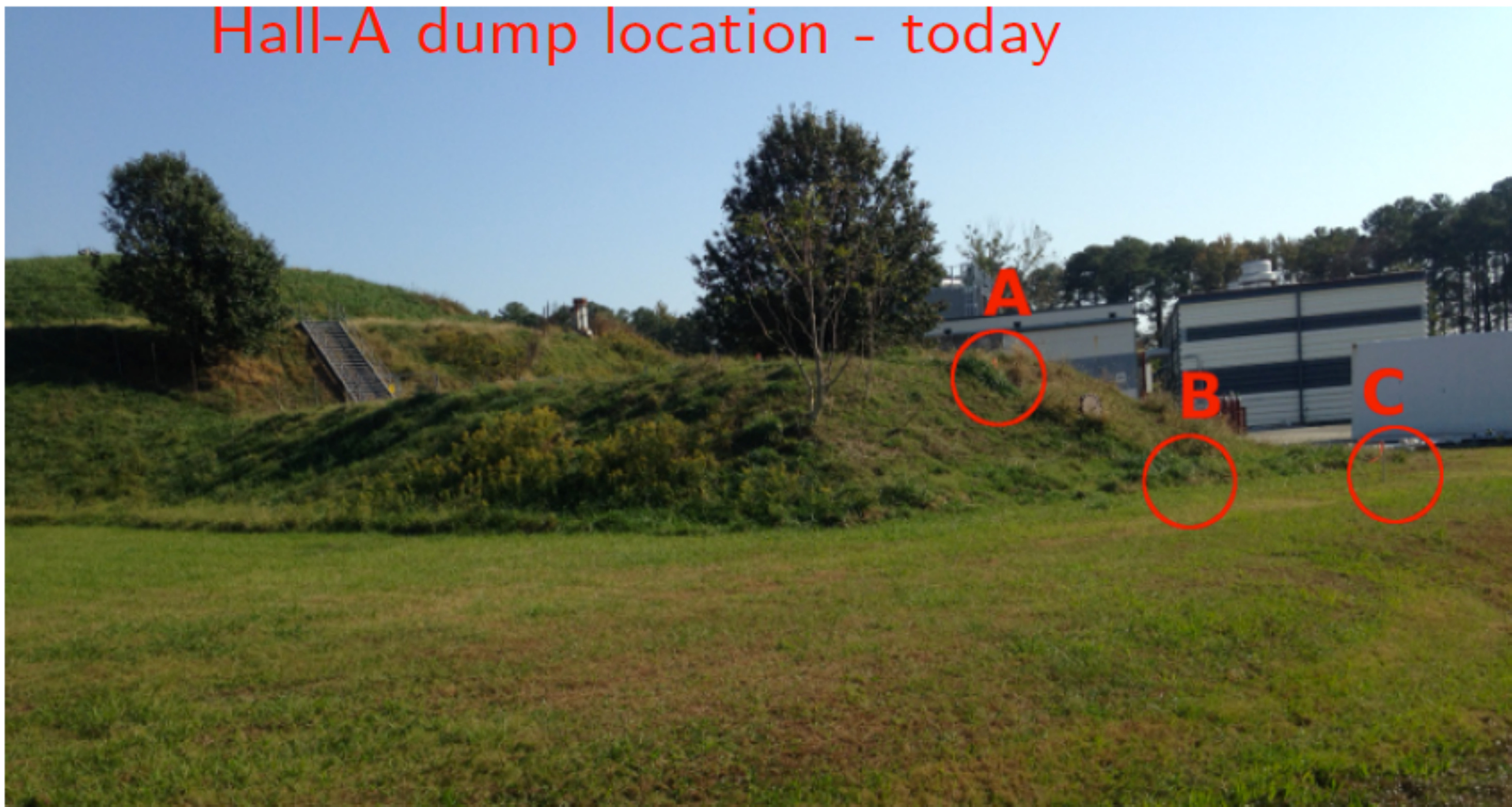}     
\includegraphics[width=11.5cm]{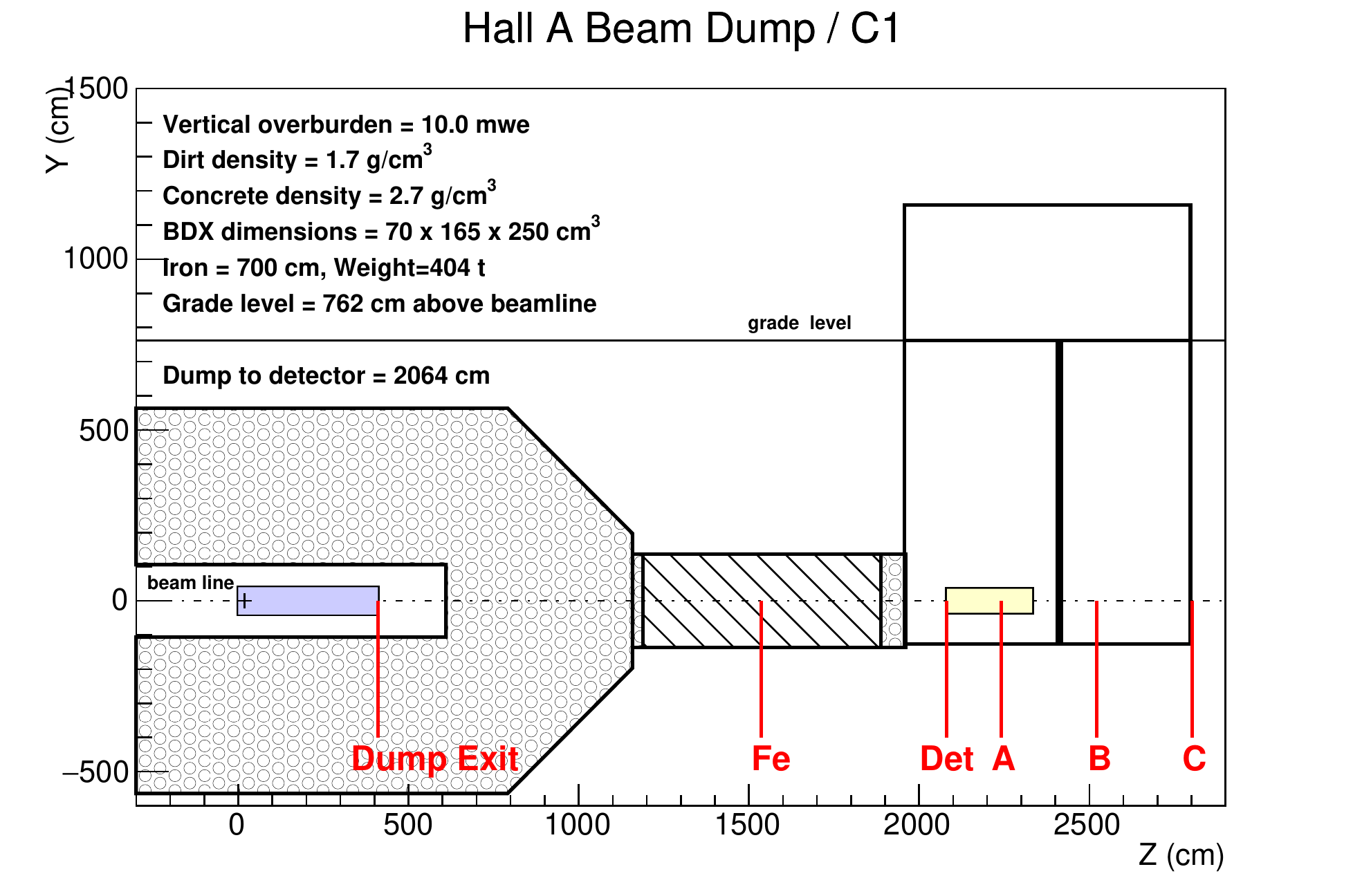}   
\caption{The area downstream of the Hall-A beam-dump and the studied test locations. Red tags show the location referred in the text. From left to right: beam-dump  exit, iron shielding, BDX detector  front face, pipes A, B and C.
\label{fig:ds-area}}
\end{figure} 
 
\subsection{Location of measurement pipes }
The area downstream of Hall-A beam-dump is shown in Fig.~\ref{fig:ds-area}  indicating test measurement locations relative to the 
new  underground facility proposed in PR12-16-001~\cite{bdx-proposal}. The three positions, indicated with markers {\bf A, B}  and {\bf C},  correspond to the hall entrance (22.4 m downstream of the beam-dump entrance), a point in the middle  (25.2 m) and the exit (28 m), respectively. The experimental set-up we are proposing requires digging a well and inserting a pipe in one (or more) of these locations. The BDX-Hodo detector (see below) will be lowered in the pipe and the muon flux sampled at different heights with respect to nominal beam height. The muon flux profiles in Y (vertical direction), measured in  different locations in Z (distance from the dump), will allow us to compare the absolute and relative MC predictions. 

\subsection{The BDX-Hodo detector \label{sec:BDX-hodo}}
The detector intended to measure the beam-on-related muon radiation and the background in the proximity of the new BDX underground facility will  make use of a BDX ECal CsI(Tl) crystal, identical to the ones proposed for the full experiment, sandwiched between a set of segmented plastic scintillators.
\begin{figure}[ht!] 
\center 
\includegraphics[width=3.3cm]{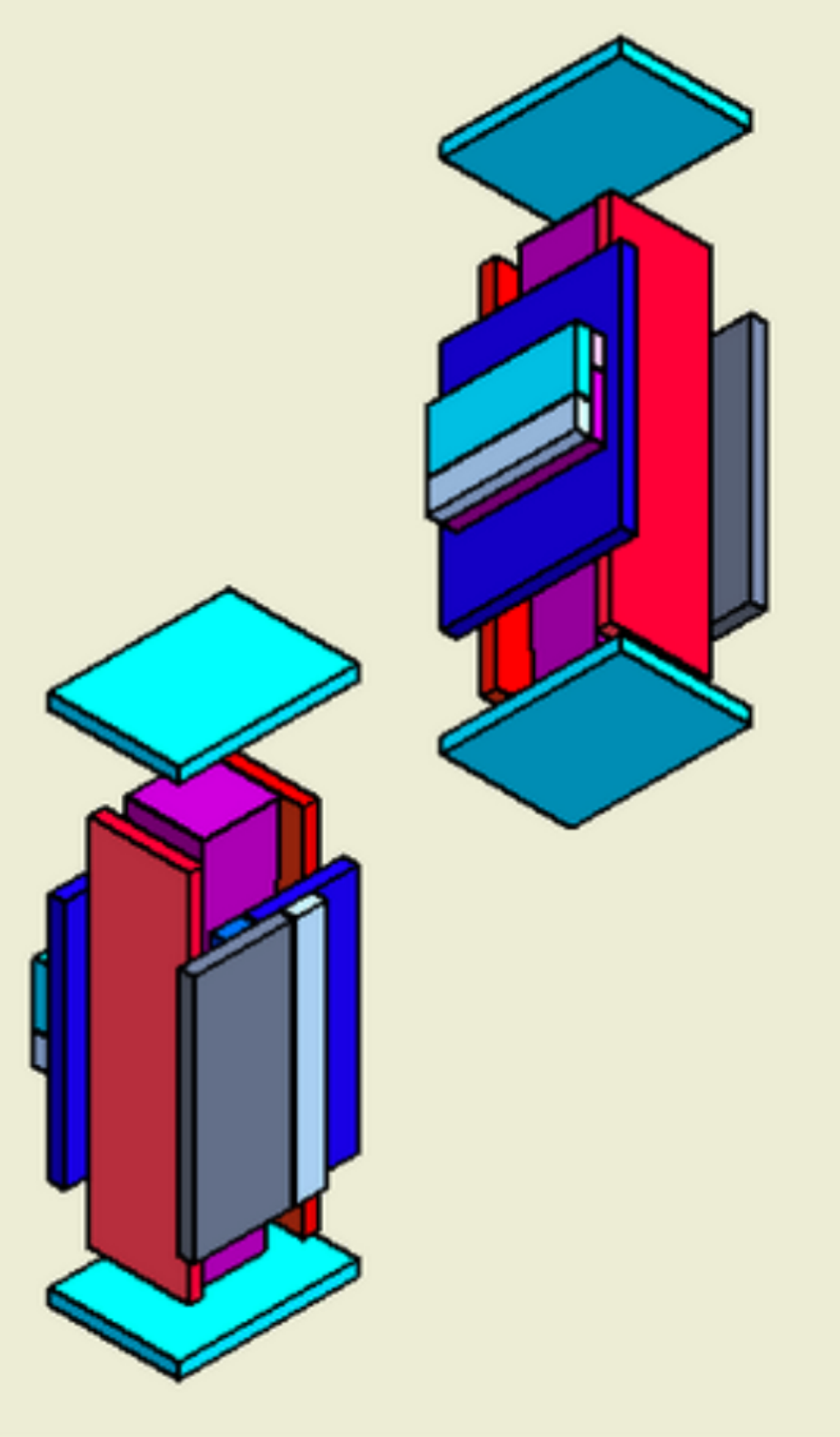}  
\caption{The CAD representation of the BDX-Hodo detector.
}
\label{fig:det-cad}
\end{figure}
The detector is assembled with technologies proposed for use in the final experiment so it will have similar sensitivities to background.
The requirement of a hit in both front and back paddles defines a 3x3 matrix of 2.5x2.5 cm$^2$ pixels providing a cm-like muon XY position resolution. 
Four more  paddles covering  the left/right sides and the top/bottom of the crystal will be used to veto cosmic rays and other radiation not associated to the beam direction.
 The scintillator paddles will be  made with clear plastic, each  read out via a WLS fiber coupled to a 3x3 mm$^2$ Hamamatsu S12572-100 SiPM sharing the same technology used in the BDX Inner Veto detector (described in details in Sec. 3.2.2 of  PR12-16-001).
The detector will be contained  in a 20-cm diameter stainless-steel cylindrical vessel, covered on top and on the bottom by steel lids. The whole assembly will  be water-tight to prevent any water from leaking inside the vessel. A stainless-steel  extension on the top cover will be used  to run  cables (signal and power)  from the detector to the ground-level. 
A loose condition  on the CsI(Tl) crystal will trigger the DAQ to record signals from all SiPMs.
Off-line, muons produced by the electron beam will be identified by requiring a 5-fold coincidence (two front paddles + CsI(Tl) crystal + two back  paddles).
The full DAQ system (crate + pc)  will be shielded in a van parked close to the well entrance. The power will be provided by a diesel power generator to minimize the requirements of long extension cords.


FLUKA has been used to generate and propagate muons to the location of interest. The  detector geometry and the realistic response of the CsI(Tl) crystal and plastic scintillators have been implemented in GEANT4 (see  Appendix B.2 of PR12-16-GEANT4~\cite{bdx-proposal} for details about the detectors response  parametrization). 
To estimate rates, we assumed a detection threshold of 10 photoelectrons in scintillators and 100 photoelectrons in the crystal corresponding to 400 keV and 2 MeV of deposited energy respectively\footnote{MIPs release $\sim50$ phe ($\sim$2 MeV) and 1670 phe ($\sim$32 MeV)  respectively.}. 

\begin{figure}[ht!] 
\center
\includegraphics[width=5.4cm]{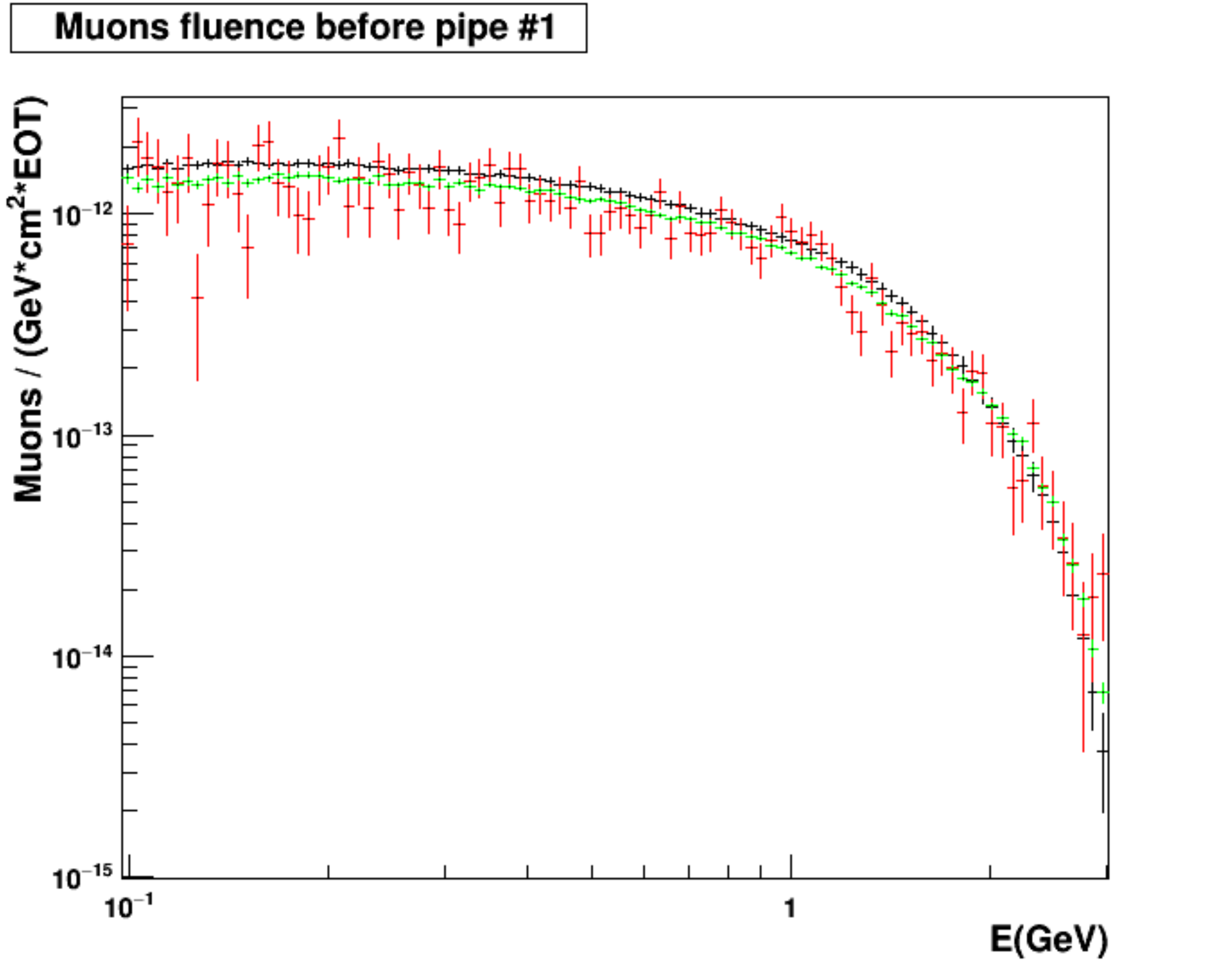}
\includegraphics[width=5.4cm]{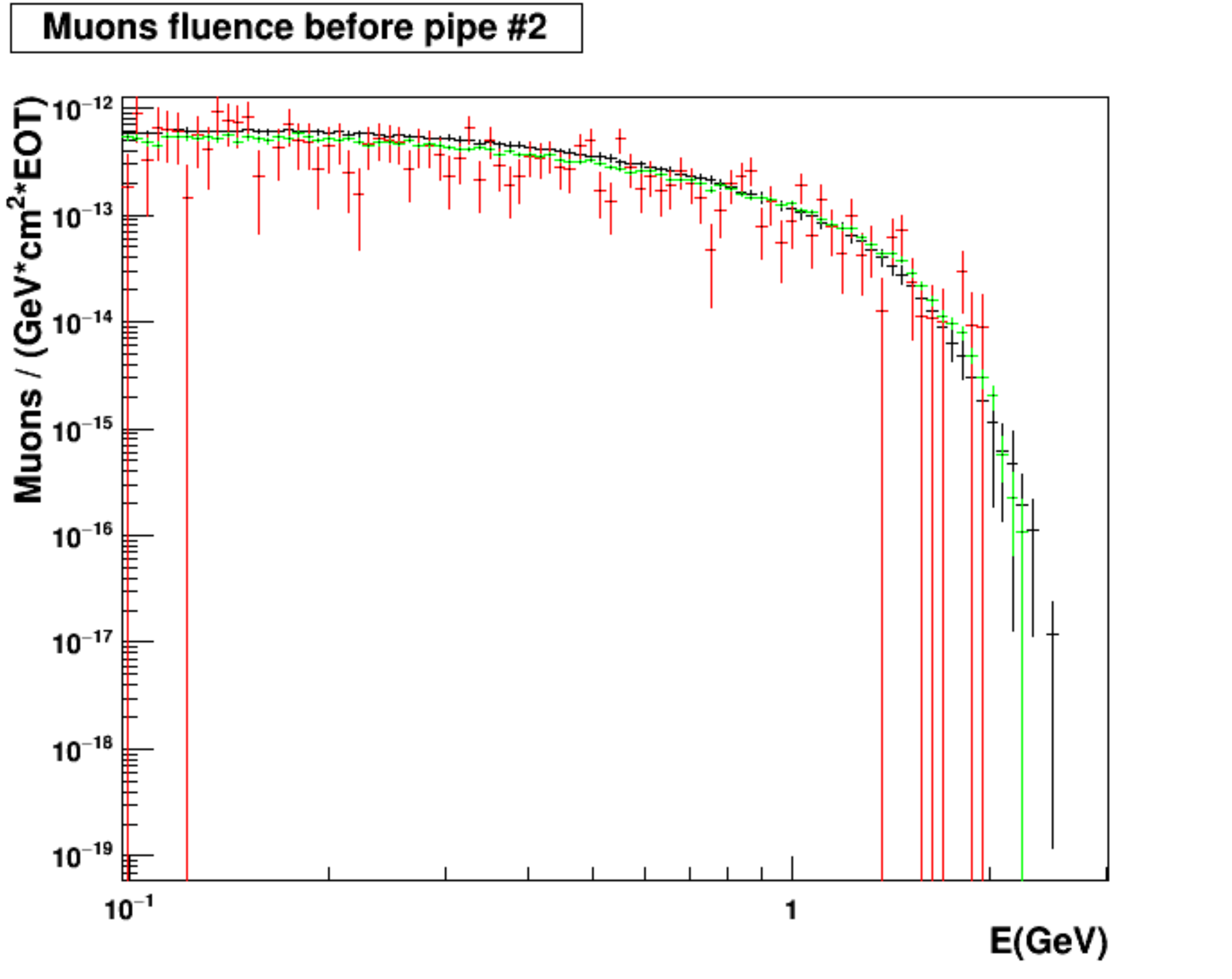}
\includegraphics[width=5.4cm]{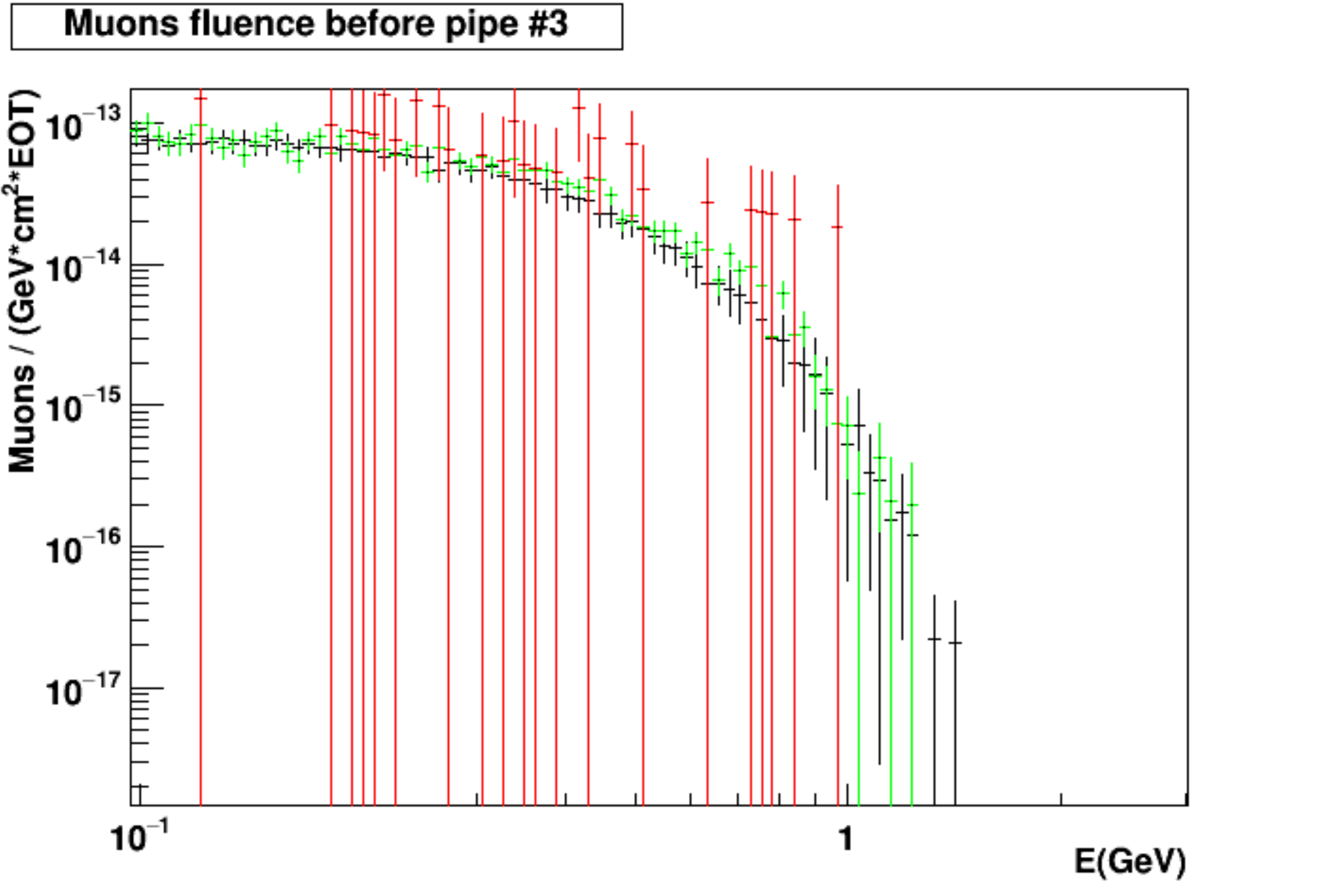} 
\caption{Differential fluence of muons at the three locations of interest (A, B and C). Beam dump interactions using  FLUKA (black), GEANT4 (red). The agreement between FLUKA and the high statistics GEANT4 calculations is very good. The statistical limitations of the GEANT4 data at location C is also apparent.}
\label{fig:mu-comp}
\end{figure}

\subsection{Expected test results} 
\subsubsection{Muon flux}
Fig.~\ref{fig:mu-comp} shows the muon flux crossing the BDX-Hodo  as obtained by GEANT4 and FLUKA  in the  three locations of interest ({\bf A}, {\bf B} and {\bf C}).
The flux has been sampled for the case when the BDX-Hodo is centered on the beamline.
Results are reported for FLUKA (black), and full GEANT4 simulations (red).
The number of events generated at the dump correspond to  (1.2 $\pm$0.1) 10$^{12}$ EOT or one second of 0.2 $\mu$A current.
Rates in crystal, scintillators and  requiring a 5-fold coincidence of the two front/back layers of plastic with  the crystal are reported in Table~\ref{tab:rate} for a beam current of 10$\mu$A and detection thresholds as listed in the previous Section. Results show a drop in rate by about one order of magnitude when moving from one location to the next. 
Table~\ref{tab:rate-height}  shows the expected rate measured in  position {\bf C}  when the BDX-Hodo detector is off-axis by 40 and 80 cm. The measurement of muon rate at different heights (angles) wrt to the beam-line (beam-dump) will provide further information to validate simulations.  
Fluxes in position {\bf C} (or/and {\bf B}) are large enough to be detectable (significantly higher than cosmic muons and beam-dump neutron background) and manageable  by crystal, SiPMs and front-end electronics (no pile-up effects expected). These two locations are close to the paved road and easily accessible by the drilling machine and related equipment. Similar conclusions (scaling rates by 10) hold  if the beam current drops/increases by one order of magnitude (between 1 and 100 $\mu$A) making the test feasible in parallel to any 11 GeV operation of Hall-A.

\begin{table}[htp]
\caption{Beam-on muon rates expected in BDX-Hodo for I$_{beam}$ = 10 $\mu A$ in locations {\bf A}, {\bf B},  and {\bf C} at beam height.}
\begin{center}
\begin{tabular}{|c|c|c|c|c|}
\hline
Location & Rate$_{Crystal}$  (kHz)&  Rate$_{upstream \;Scint} $(kHz) & Rate$_{Coin}$ (kHz) \\
\hline\hline
{\bf A}  & 120  &120   & 24 \\
 \hline
{\bf B} & 20 &  22 & 3.7 \\
 \hline
{\bf C} & 2.8 &  2.5 & 0.5 \\
\hline\hline
\end{tabular}
\end{center}
\label{tab:rate}
\end{table}

\begin{table}[htp]
\caption{Beam-on muon rates expected in BDX-Hodo for I$_{beam}$ = 10 $\mu A$ in position {\bf C} sampled at three vertical distances from the beam-line.}
\begin{center}
\begin{tabular}{|c|c|c|c|c|}
\hline
Vertical distance  & Rate$_{Crystal}$  (kHz)&  Rate$_{upstream \;Scint} $(kHz) & Rate$_{Coin}$ (kHz) \\
\hline\hline
0 (nominal)  & 2.8  &2.5    & 0.5 \\
 \hline
40 cm & 1.4 &  2.5& 0.17  \\
 \hline
80 cm&  0.6  & 0.6   &  0.08  \\
\hline\hline
\end{tabular}
\end{center}
\label{tab:rate-height}
\end{table}
For the sake of completeness, the muon flux has also been evaluated using FLUKA in the closest locations above ground illuminated by the  beam-dump vault.  Possible measurements
with detector located above ground so no drilling is required, and thereby simplify the test, would expect no practical signal. Integrating  over the surface of the BDX-Hodo detector ($\sim 100$ cm$^2$) 
and considering as a reference a beam current of 10 $\mu$A, no sizable muon flux would be detected (Rate$_{Max}<$3 Hz). 

\subsubsection{Other beam-related backgrounds}
\begin{figure}[ht!] 
\center
\includegraphics[width=5.42cm]{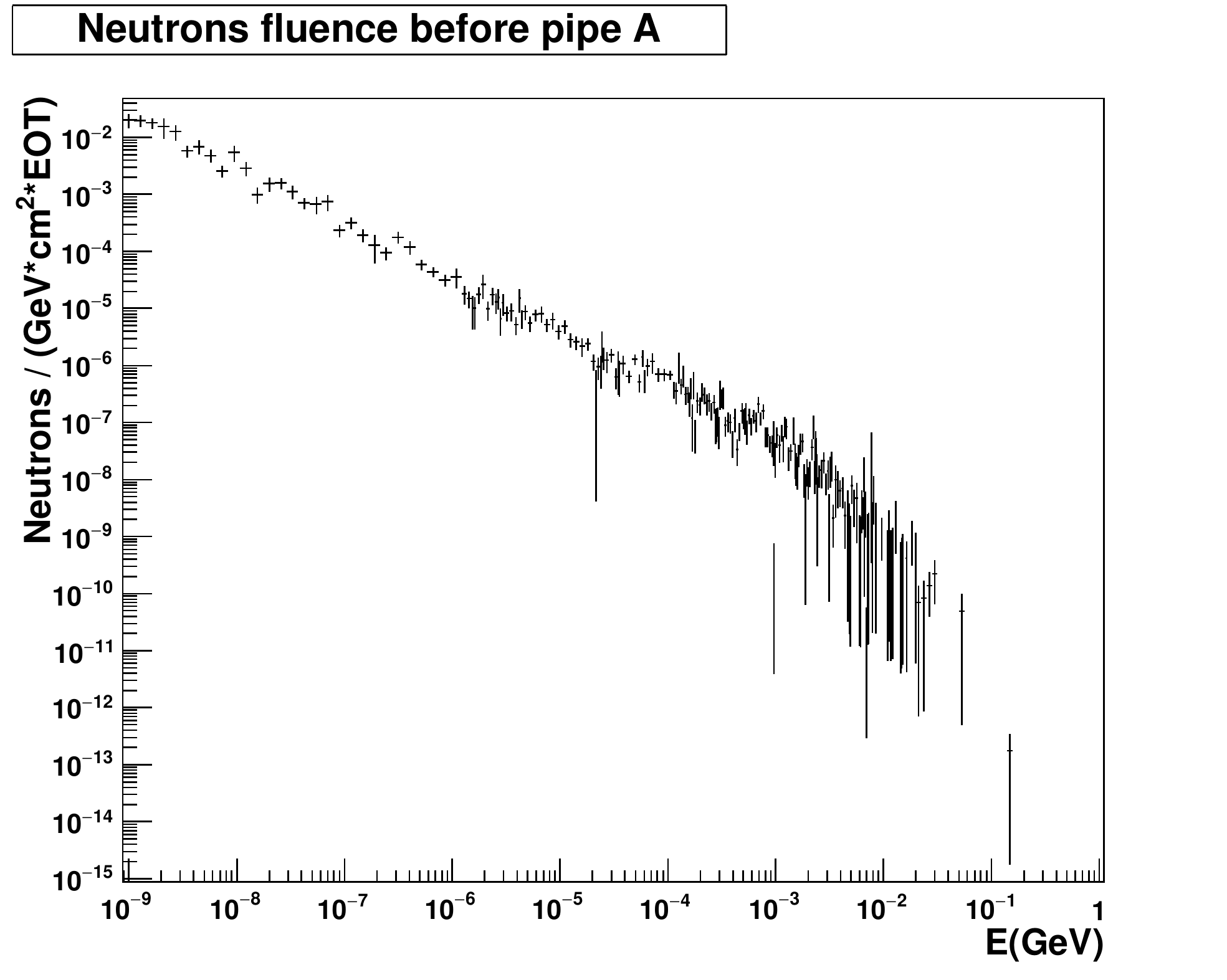} 
\includegraphics[width=5.42cm]{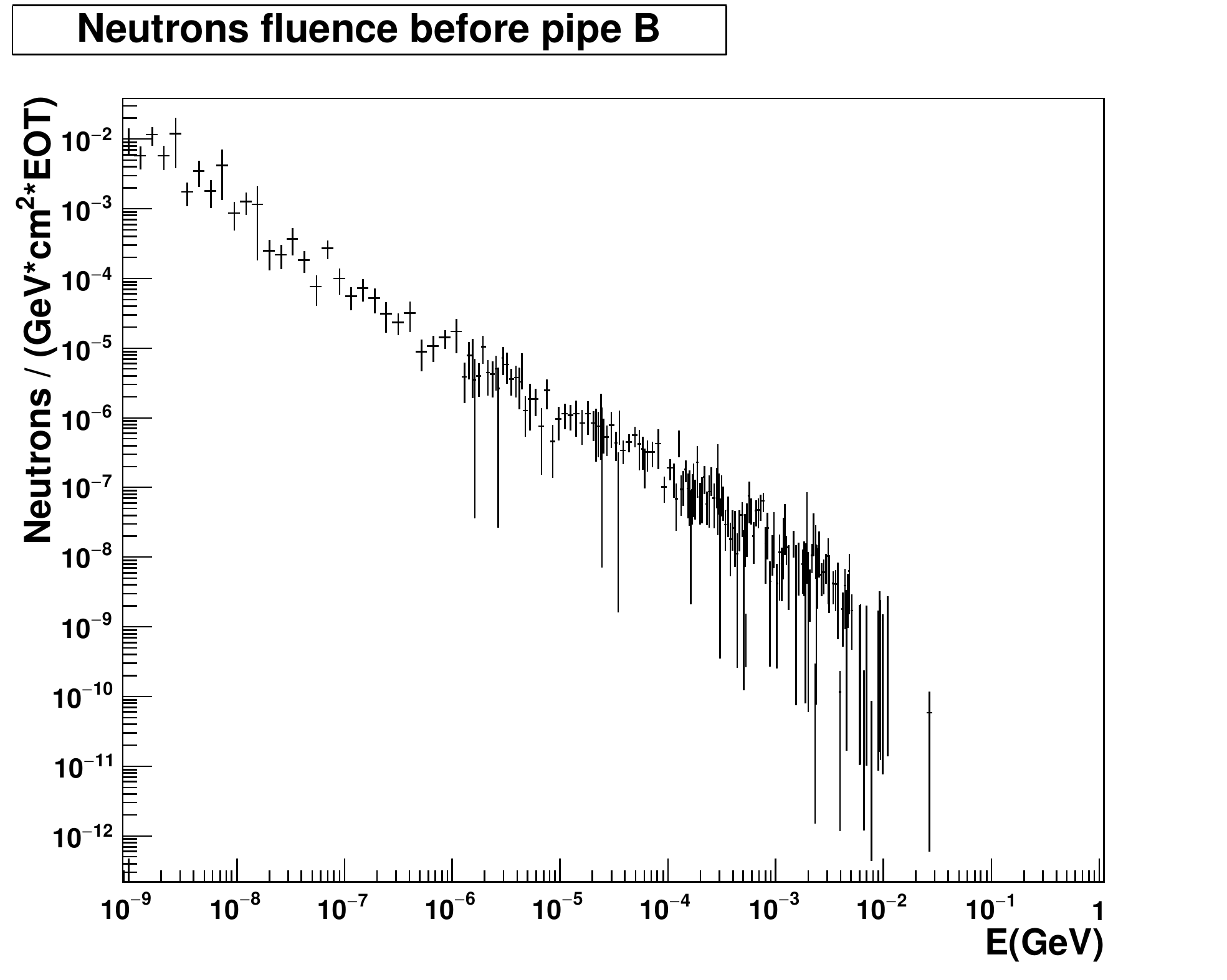}
\includegraphics[width=5.42cm]{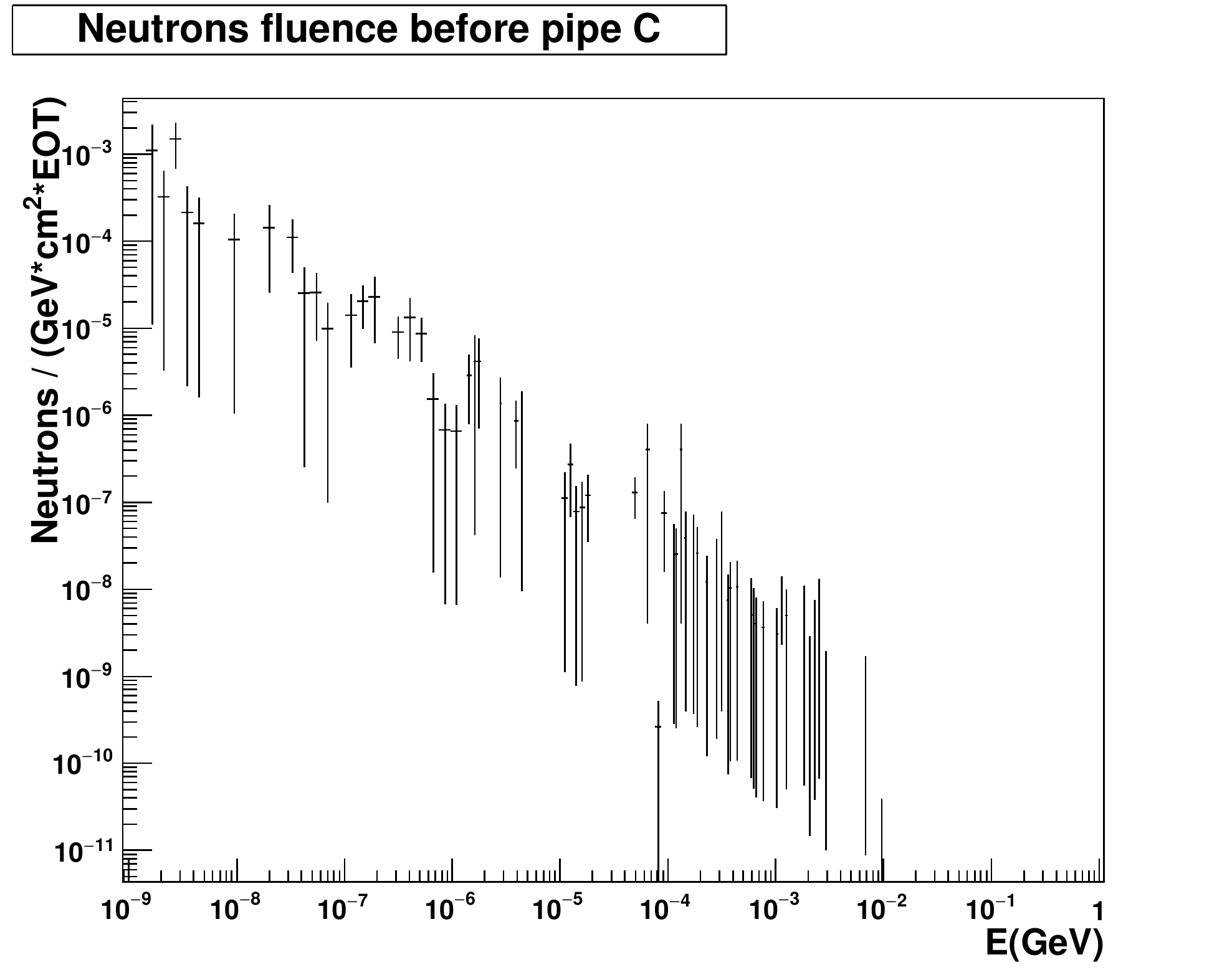}
\caption {Neutron differential fluence at the three locations of interest. Spectra are obtained from electron beam interaction with the beam-dump using FLUKA.}
\label{fig:nu-comp}
\end{figure}

Beside muons, other particles are produced in the 11 GeV electron beam interaction with the dump.
The majority (electrons, gamma, nuclei and fragments) are ranged-out well before to reach the region of interest but some (low energy neutrons, mainly) may propagate through  concrete and  dirt reaching the BDX-Hodo detector.
Fig.~\ref{fig:nu-comp} shows the neutron flux  in the  three locations of interest as obtained by  FLUKA simulations. 
\begin{figure}[ht!] 
\center
\includegraphics[width=4.9cm]{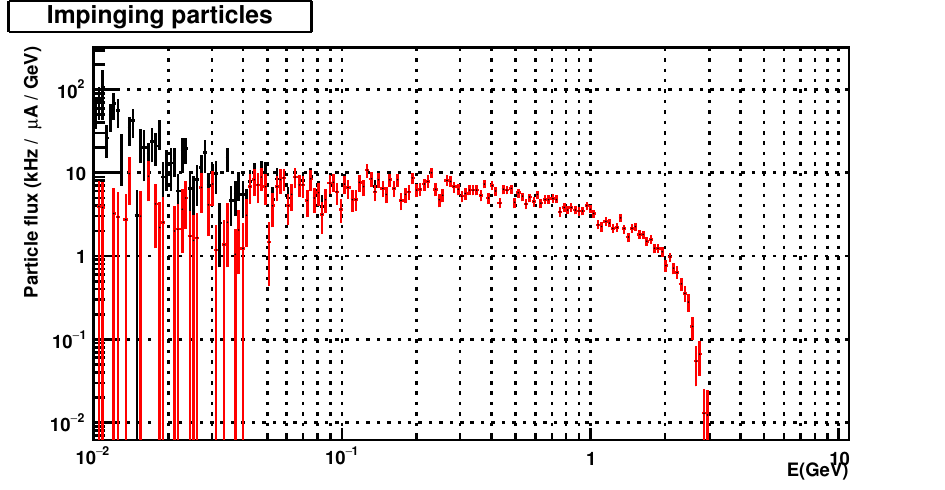}
\includegraphics[width=5.6cm]{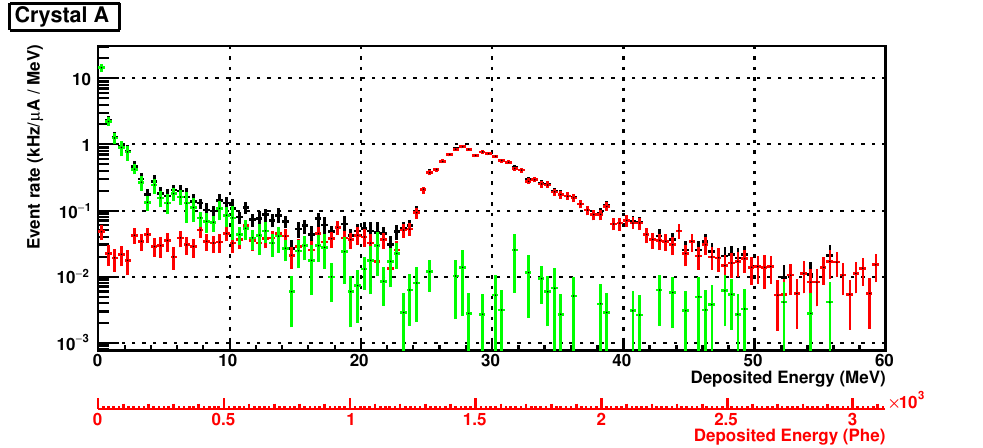}
\includegraphics[width=5.6cm]{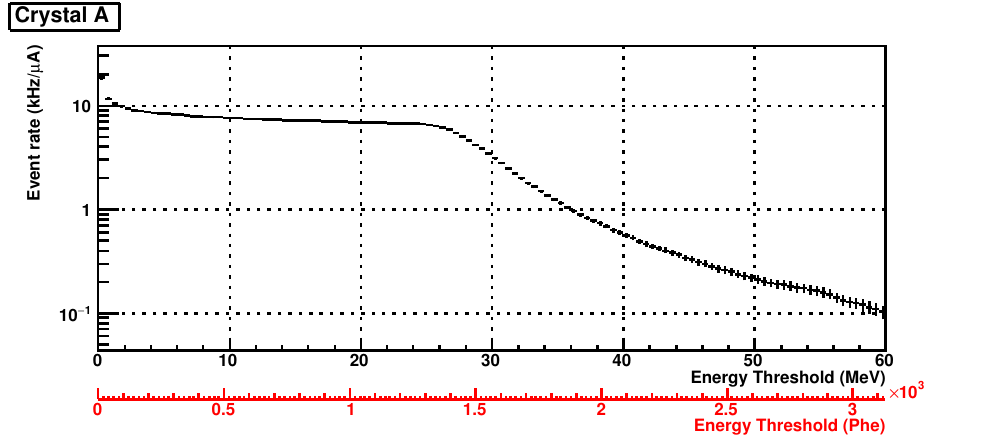}  
\includegraphics[width=4.9cm]{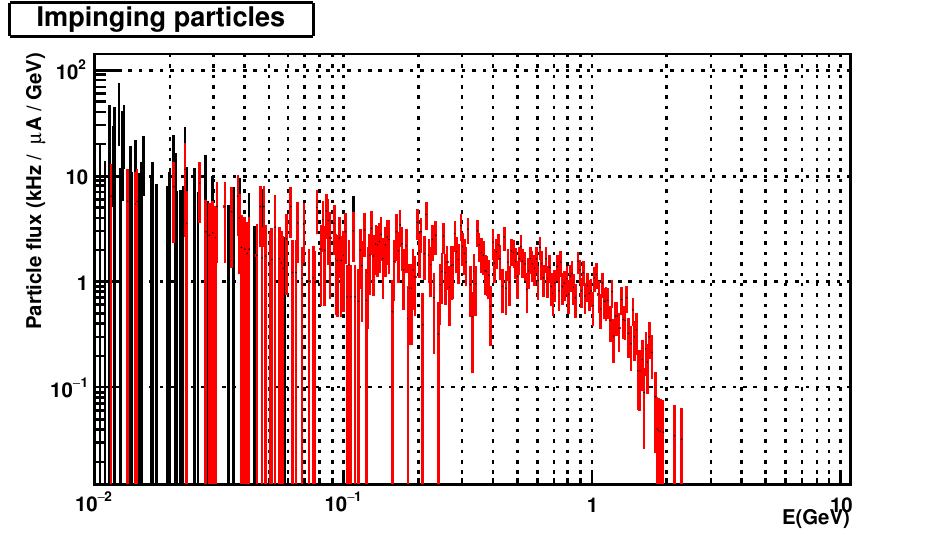}
\includegraphics[width=5.6cm]{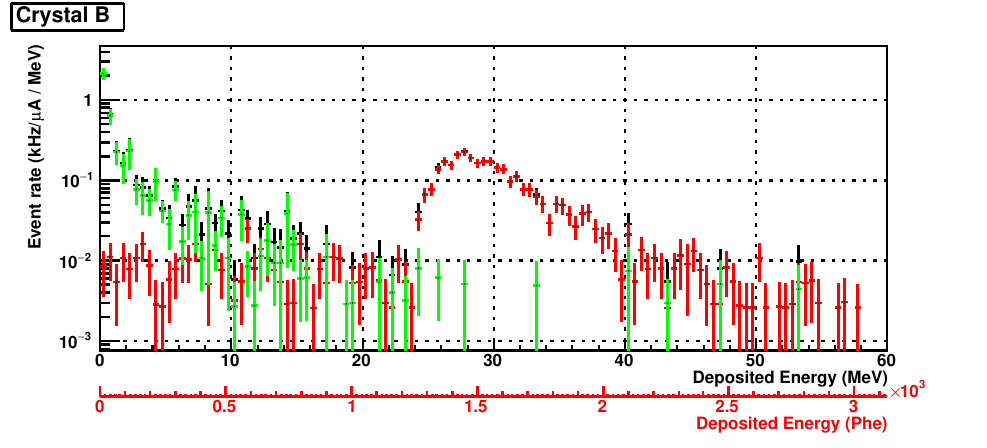}
\includegraphics[width=5.6cm]{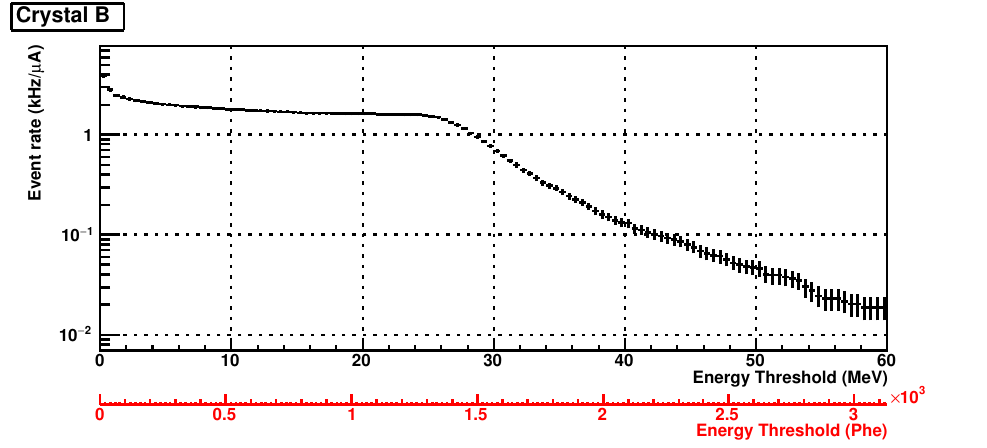}  
\includegraphics[width=4.9cm]{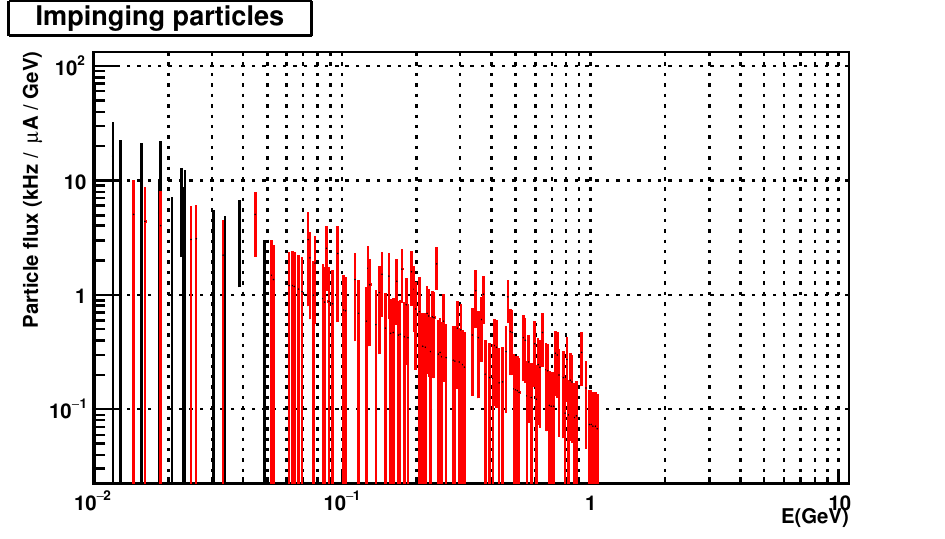}
\includegraphics[width=5.6cm]{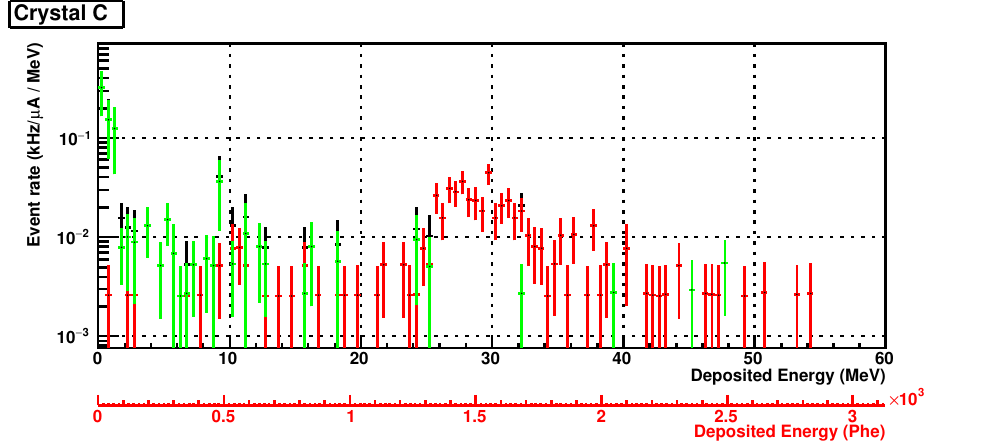}
\includegraphics[width=5.6cm]{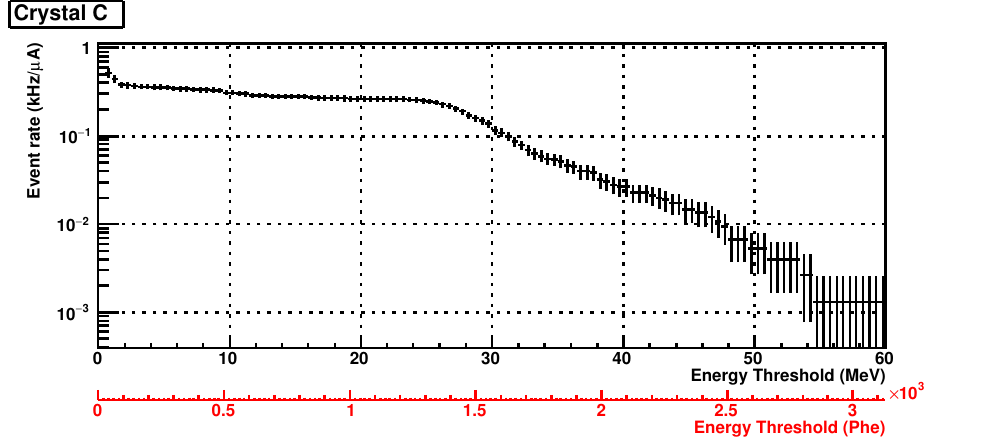}  
\caption {Left: fluence  of all particles (black) and muons only (red) hitting the CsI(Tl) crystal. Middle: distribution of energy deposited in the CsI(Tl) crystal by all particles (black), muons only (red) and the rest (green). Right: integrated rate as function of deposited energy.
Plots refer to the crystal located in {\bf A} (top),  {\bf B} (middle), and {\bf C} (bottom).}
\label{fig:fluence}
\end{figure}

For a complete understanding of the low energy ($<$ MeV) background in the BDX-Hodo crystal, particles produced in the dump cannot be tracked separately since some of them are produced along the way (e.g by energetic muons or neutrons in the proximity of the detector). On top of that, neutral particles (in particular low energy/thermal neutron) do not directly interact with the crystal but deposit a visible energy via secondary interactions (e.g. gamma from nuclear capture in the surrounding material) making hard, if not impossible,  to track back the source in the dump. For all the above mentioned reasons we evaluated the background by running the  full FLUKA simulation of  11 GeV electrons interaction with the beam-dump recording  the deposited energy in BDX-Hodo crystal. 
Figure~\ref{fig:fluence}-left shows the fluence of all particles (black) and muons only (red) on the CsI(Tl) surface. As already noticed, the high energy range of the spectrum is saturated by muons. 
Figure~\ref{fig:fluence}-middle shows the deposited energy in the CsI(Tl) crystal (located in the three  position of interest).  Muons (shown in red) almost saturate the highest energies (the MIP peak is clearly visible around E$_{Dep}$ = 32 MeV) while the contribution from other particles (neutrons) accumulates at low energies.

The crystal  integrated rate is reported in Fig.~\ref{fig:fluence}-right  as a function of the deposited energy (and detected photoelectrons). Considering that  the experiment only records events 
with a deposited energy in the crystal  larger than  1 MeV (25 pe), the expected rate is in the range of 10 kHz.

\subsubsection{Cosmic rates}
The cosmic muon background in the BDX-Hodo has been evaluated using GEANT4. This is the same cosmic flux generator used in PR12-16-001~\cite{bdx-proposal}. The muon energy spectrum has been divided in different ranges  and correctly weighted to estimate the full rate expected on the detector. Rates have been evaluated for CsI(Tl) crystal, Top scintillator,  and for the coincidence of the front/back scintillator with the crystal  to mimic the condition used to identify and account for beam-on muons. We assumed the same detection thresholds used in the other rate estimates (10 phe and 100 phe for scintillators and crystal respectively). Tab.~\ref{tab:cosmic} shows the results of this study. The cosmic muon  rate is negligible (in every condition $<$ 1 Hz) well below the expected rate of muons from the electron beam interacting in the beam dump.

The same procedure was used to generate and sample cosmic neutrons. The corresponding rate in the CsI(Tl) crystals of hits over threshold  were found to be negligible ($<$ 0.1 Hz).\
We expect  a negligible  environmental background contribution to the detected counting rates.  Cosmic and environmental background will be assessed tacking data with the beam off.

\begin{table}[htp]
\caption{Cosmic rate expected in different components of BDX-Hodo}
\begin{center}
\begin{tabular}{|c|c|c|c|}
\hline
Energy range  (GeV) & Rate$_{Crystal}$  (Hz)&  Rate$_{Top\,Scintillator} $(Hz) & Rate$_{Coincidence}$ (Hz) \\
\hline\hline
 0.2 - 2  & 0.01 &  0.02 & 0\\
 \hline
 2 - 10  & 0.2 &  0.25 & 0.01\\
 \hline
 10 - 100  & 0.35 &  0.4 & 0.01\\
\hline\hline
Cosmic muon rate  & 0.56 &  0.67 & 0.02 \\
\hline\hline
\end{tabular}
\end{center}
\label{tab:cosmic}
\end{table}%

\subsection{Costs, work- \& time-plan}
A detailed costs list and work/time-plan have been reported in  the Appendix of~\cite{mutest-note} and presented to JLab management. Here we can just mention that, after drilling two wells in   locations {\bf B} and {\bf C} in the area downstream of the Hall-A beam-dump, the tests are expected to last  approximately  4 calendar days during any time that 11-GeV beam with relatively steady current (between 1 and 100$\mu$A) is delivered  to Hall A. If possible  we would like to take data at more than one beam current to check that the count rates scale.
This  would require 1h of dedicated beam-time  coordinated with the Hall-A physics program to change the beam current by a factor of 10. Since the pipes will remain in place, it is worth noting that it will be possible to plan other opportunistic measurements with different Hall-A beam current/energy  set-ups.

\subsection{Summary of test measurement} 
To validate MC tools and gain confidence in the beam-on background shielding optimization  for the  BDX experiment we propose to measure the  muon flux in the region where the new underground facility will be located. 
We itemize the proposed test set up and the expected results:
\begin{itemize}
\item{muons produced in the dump can be measured by placing a  detector downstream of the dump at beam height, i.e. below ground;}
\item{a detector  (BDX-Hodo) based on one CsI(Tl) crystal from  the BDX ECal, sandwiched between layers of scintillator counters  will be specifically built for this measurement;}
\item{ two wells equipped with 10' pipes will be drilled  25.2 m and 28 m downstream of the beam-dump and the BDX-Hodo detector will lowered into a pipe down to beam height;}
\item{rates of beam-on muons measured by BDX-Hodo are expected  to be sizeable for  a beam current of 10 $\mu A$ ($\sim$3 .7kHz at  25.2 m (configuration B) and and $\sim$0.5 kHz at 28 m (configuration C) downstream of the dump);}
\item{given the count rates reported above, this measurement could be done  with a variety of beam current (1 - 100 $\mu A$ ) making the test fully parasitic wrt the Hall-A plans;}
\item{this measurement was found to be insensitive to the cosmic muon background and other backgrounds (mainly) neutrons generated in the dump; }
\item{the use of a BDX CsI(Tl) crystal will validate the proposed technology in a background-rich environment without the additional shielding needed for the BDX experiment\footnote{The BDX experiment foresees  an optimized shielding that will drastically reduce  any possible background.}; indeed, we'll be able to test the DAQ performance in a realistic configuration, and demonstrate the insensitivity to the low-energy neutron-induced pile-up on SiPM.}
\item{once the pipes are installed, tests will run for about a week, parasitically  to  any 11 GeV 1-100 $\mu$A, Hall-A run;} 
This  test, measuring the muon flux (absolute and relative) at two locations in Z (distance from the dump) and several in Y (vertical) will address the concern expressed by PAC44 report about the beam-on backgrounds in the BDX experiment.
\end{itemize}
